\documentclass[aps,prx,showpacs,twocolumn,reprint,superscriptaddress]{revtex4-2}

\usepackage{qcircuit}
\usepackage[dvips]{graphicx}
\usepackage{amsmath,amssymb,amsthm,mathrsfs,amsfonts,dsfont,mathtools,physics}
\usepackage{epsfig}
\usepackage{braket}
\usepackage{hyperref}
\hypersetup{
    colorlinks=true,
    linkcolor=teal,
    filecolor=magenta,      
    urlcolor=violet,
    citecolor=violet
}
\usepackage{bm}
\usepackage{enumerate}
\usepackage{color}
\usepackage{graphicx}
\usepackage{textcomp}

\usepackage{enumitem}
\usepackage[capitalize]{cleveref}
\usepackage[normalem]{ulem}
\usepackage{comment}
\usepackage{xcolor}

\usepackage{amsthm}

\Crefname{theorem}{Theorem}{Theorems}
\theoremstyle{remark}

\newcommand{\sqrtswap}{{\sqrt{\text{SWAP}}}}

\newcommand{\qmaddress}{\affiliation{Quantum Motion, 9 Sterling Way, London N7 9HJ, United Kingdom}}
\newcommand{\oxddress}{\affiliation{Department of Materials, University of Oxford, Parks Road, Oxford OX1 3PH, United Kingdom}}

\begin{document}

\title{{\em Ab initio} modelling of quantum dot qubits: \\ Coupling, gate dynamics and robustness versus charge noise}
\author{Hamza Jnane}
\email[hamza.jnane@materials.ox.ac.uk]{}
\qmaddress
\oxddress

\author{Simon C. Benjamin}
\qmaddress
\oxddress

\begin{abstract}
    Electron spins in semiconductor devices are highly promising building blocks for quantum processors (QPs). Commercial semiconductor foundries can create QPs using the same processes employed for conventional chips, once the QP design is suitably specified. There is a vast accessible design space; to identify the most promising options for fabrication, one requires predictive modelling of interacting electrons in real geometries and complex non-ideal environments.
    In this work we explore a modelling method based on real-space grids, an {\em ab initio} approach without assumptions relating to device topology and therefore with wide applicability. Given an electrode geometry, we determine the exchange coupling between quantum dot qubits, and model the full evolution of a $\sqrtswap$ gate to predict qubit loss and infidelity rates for various voltage profiles. We determine full, 3D solutions and introduce a method which can obtain near-identical predictions using far more efficient 2D computations. Moreover we explore the impact of unwanted charge defects (static and dynamic) in the environment, and test robust pulse sequences. As an example we exhibit a sequence  correcting both systematic errors and (unknown) charge defects, observing an order of magnitude boost in fidelity. The technique can thus identify the most promising device designs for fabrication, as well as bespoke control sequences for each such device.
\end{abstract}

\maketitle

\section{Introduction}
Electron spins in gate-defined semiconductor quantum dots \cite{burkard_2023} are a strong candidate for qubit implementation \cite{mills_2022, xue_2022, noiri_2022}. Due to their small size, fast operations, and compatibility with current industry standards \cite{de_michielis_2023, zwerver_2022, veldhorst_2017, maurand_2016} silicon spin qubits hold a great promise for the development of full scale, mature-era quantum computers.

The vast design space provided by semiconductor foundries is a highly enabling, while also presenting a challenge to identify the most promising device architecture. The iterative nature of chip development, coupled with the time and cost of a chip design cycle, makes it highly desirable to have accurate predictive modelling tools to speed up the discovery of scalable designs. Given a candidate chip layout defined using standard commercial software (as employed for traditional chips), such tools would extract key metrics and enable one to optimise the design before committing to fabrication.

The electron spin qubits are confined by potentials generated by the device's electrodes, and the design space for electrode layout (and corresponding voltage ranges) is wide. Of particular interest are double dot potentials where two electrons can be isolated and experience exchange coupling: the spin-spin interaction resulting from underlying charge state symmetries. The strength $J$ and its controllability ultimately limit the speed and quality of any native two-qubit gate \cite{loss_1998, divincenzo_2000, meunier_2011}. Naturally, there is a long-standing effort to predict the exchange coupling for a given double dot potential (and hence, for a given electrode layout) by either analytic or numerical methods~\cite{saraiva_2007, anderson_2022, ercan_2021, calderon_2006}. The task is complex and typically simplifications to the device layout and the environment are adopted in order to make the task possible. 

A key consideration in optimising the fidelity of two-qubit gates is the sensitivity of the exchange coupling to  charge noise \cite{reed_2016, culcer_2009, nielsen_2010}. 
Accounting for the qubits' environment is thus necessary for a predictive tool to be practically useful. Recently, several papers have addressed this challenge~\cite{cifuentes_2023, buonacorsi_2020, shehata_2022} and computed the exchange coupling for realistic scenarios. Authors have adopted techniques to make the otherwise-costly numerics more tractable; for example by tailoring the basis so as to enable a compact form for the Coulomb interaction, by coarse-gaining the model in some direction(s), or by reexpressing the problem through a path integral formulation.  Ref.~\cite{buonacorsi_2020} limited the inclusion of charge noise to optimising devices geometries to reduce the sensitivity of $J$ to electrical fluctuations, while \cite{cifuentes_2023, shehata_2022} explicitly added charge noise to their model using a two-level fluctuator model. Recent experiments \cite{elsayed_2022} confirmed the validity of this choice. 

The approach taken in the present paper is {\it ab initio} in the sense that it makes very few assumptions regarding the anticipated physics but sees the key properties emerge once suitable parameters are found. We use a uniform real-space grid method which described the relevant two-electron states for {\it any} reasonably smooth potential landscape without requiring that it be approximately locally harmonic, or that eigenstates be close to any given analytic form, etc. Since we do not use the low-lying eigenstates of the particle(s) as our means of representation, we are not constrained to low-energy dynamics. The representation fails only for states more highly curved than the resolution of the grid basis -- over a billion elements in some cases. Correspondingly the numerical modelling is fairly demanding, but can tackle quite general scenarios.  For the present study we focus on a flat-bottomed double well potentials derived, via a Poisson solver, from a realistic device's electrode geometry. We begin by identifying a different $J$-coupling scenarios including a suitable `off' configuration, where the notion of `left' and `right' qubits emerges as an excellent approximation to the true state. 

We directly model the dynamics of a $\sqrtswap$ gate as the confining potential continuously deforms, propagating the full two-electron state forward with a split-operator (SO) method. At gate completion we reassess the validity of the qubit representation, i.e.  whether there is increase in the $(2,0)$ \& $(0,2)$ configurations which correspond to qubit loss, and we report this separately from the fidelity of the gate operation within the $(1,1)$ subspace. We observe the effect of non-adiabatic voltage ramp on this loss probability, and verify that a suitable ramp introduces negligible loss. We then proceed to assess the impact of unwanted trapped charges in the environment, either as static entities or as fluctuators that switch charge position during the gate operation. By extracting the behaviour of our full model into a reduced basis, we rapidly explore options for robust pulse sequences -- i.e. a multi-stage gate that corrects for intrinsic axis-error as well as charges in the environment. Adapting methods from the NMR literature \cite{levitt_1986, brown_2004, jones_2009, jjonesNMRnotes} we exhibit a 9-pulse sequence with remarkable robustness against both effects, albeit at the cost of speed. 

\begin{figure*}
    \centering
    \includegraphics[width=\textwidth]{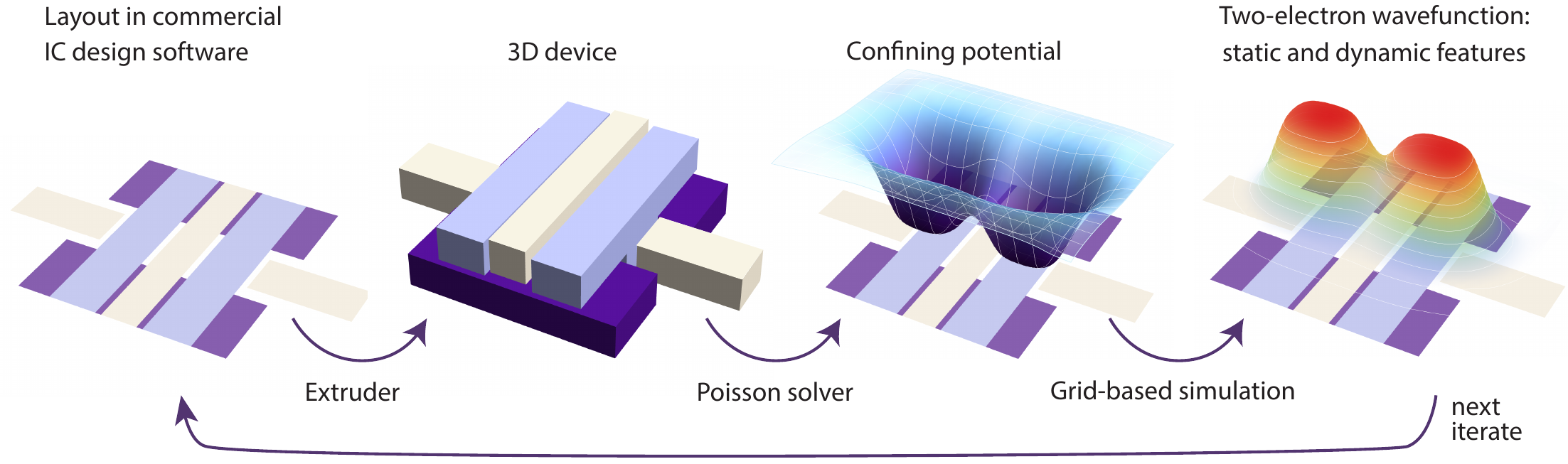}
    \caption{Schematic of the iterative design process of a quantum dot device. Starting from an idea of an integrated circuit layout we would like to know the performance of the chip without sending it through the lengthy fabrication process. Using an extruder we generate a 3D version of the chip for which we specified the different materials. This in turn allows us to extract realistic quantum dot potentials. The grid-based modelling tool presented in this paper then gives us access to the device's key performance metrics. For instance, we could discard designs that show a small exchange coupling between neighbouring dots. On top of the static properties, we can also study the qubits' dynamics and e.g. compute two-qubit gate fidelities, how they are impacted by the presence of charge noise and how we can mitigate this effect. Finally, this process will inform new layouts and through similar iterations will lead to more promising devices that can be sent for fabrication.}
    \label{fig:main_figure}
\end{figure*}

The tool we describe is capable of characterising the impact of arbitrary charge defects, within arbitrarily-changing electrode potentials and in a framework that fully describes qubit leakage. Moreover this is possible both for 2D and true 3D particle pairs. As far we are aware there is no prior-reported tool with these capabilities, although aspects have been previously explored, for example charge noise modelled as two-level fluctuators have been studied \cite{shehata_2022, cifuentes_2023}, and first-principles methods to find exchange coupling \cite{shehata_2022, buonacorsi_2020, cifuentes_2023}. A key motivation is to facilitate improved design cycle times for the fabrication of semiconductor quantum devices, since electrode configurations and potential ramping can be tested against various charge noise scenarios prior to committing to a design (see \cref{fig:main_figure}).

This paper is organised as follows. In \cref{sec: model}, we briefly explain how our double quantum dot and electrons' wavefunctions are modelled using real-space grids. Then, in \cref{sec: exchange}, we study how our tool can be used to compute the exchange coupling in an idealised environment. In \cref{sec: time_dynamics} we explore the dynamics of the exchange gate and in \cref{sec: charge_noise} we study how it is impacted by the presence of charge noise. Finally in \cref{sec: robust_pulse}, we assess the mitigation power of different pulse sequences and conclude in \cref{sec: conclusion}. 
\section{\label{sec: model} Model}

\subsection{Hamiltonian}
Here, we focus on the study of two electrons trapped in a double quantum
dot (DQD) potential (see \cref{fig:state_prep}). Under the effective mass approximation, the Hamiltonian of the system is given by,
dot (DQD) potential (see \cref{fig:state_prep}). Under the effective mass approximation, the Hamiltonian of the system is given by,

\begin{align}{\label{eq: hamiltonian}}
  H &= \frac{\hbar^2}{2m^{*}}\mathbf{p_{1}^{2}} + \frac{\hbar^2}{2m^{*}}\mathbf{p_{2}^{2}}
    + V(\mathbf{r_1})+V(\mathbf{r_2}) \nonumber \\
    &+ \frac{e^2}{4\pi\epsilon_0\epsilon_r|\mathbf{r_1}-\mathbf{r_2}|}
    + g(\mathbf{r_1})\mu_B\mathbf{B}\cdot\mathbf{S_1} + g(\mathbf{r_2})\mu_B\mathbf{B}\cdot\mathbf{S_2},
\end{align}
with $m^{*} = 0.19m_e$ the effective mass of an electron in
silicon, $\epsilon_r = 11.68$ silicon's
relative permittivity, and $g(\mathbf{r_i})$ the position-dependent electron $g$-factor. The potential $V$ is generated using a Poisson solver on realistic devices. In this first study we neglect certain additional effects, while noting that their inclusion would be entirely possible within the approach and is an attractive direction for further work. Specifically, we neglect the impact of the magnetic field on the orbital motion of the electrons (through its associated vector potential) as we are concerned with flat or nearly-flat 2D dots subject to an in-plane field. Moreover, we neglect the valley degree of freedom as the valley splitting is usually large in SiMOS devices. Finally, we also neglect spin-orbit coupling, noting that it is small in silicon. Presently we remark on the ways our tool could be extended to encompass some of these additional features.

\subsection{\label{sec: gbsim} Grid-based simulation}

\subsubsection{Introduction to grid-based and split-operator methods}

One of the canonical approaches to modelling either single or multiple quantum particles, is to represent their state by storing a complex amplitude for each point in a regular grid -- such a grid may be in real-space, $k$-space, or alternately in each representation. The approach is broadly a discrete variable representation (DVR)~\cite{DVR2000}, and the approach to driving dynamics used in this paper is a split-operator Fourier transform (SO-FT) method, which dates back at least as far as the work of Feit and Fleck in 1976~\cite{FeitAndFleck76}. Widely used in a variety of contexts, the method involves repeatedly Fourier transforming between the real-space and the dual $k$-space representations, described below. 

It is worth noting that the method has recently received attention from researchers seeking efficient algorithms for modelling many-body systems on future quantum computing systems. The use of the Fourier transform is a particularly natural fit since this transformation is well-understood and efficient on a quantum machine. Indeed, the specific variant of the SO method that we use here is one that has been well-explored in that context (see e.g. Refs.~\cite{RyanLDQSM,hans_grid_based_2023}).

\subsubsection{State representation}

It is convenient to begin in the $k$-space picture, and to start by describing a single particle in 1D.
The state of this system is represented  as, 
\[
\Psi_\textrm{KS}(x) = L^{-1/2}\sum_{k=-m}^{m-
1}c_k\exp\left( \tfrac{2i\pi k x}{L}\right)
\]
where we write the subscript KS for $k$-space. In the expressions in this paper we typically set the `box size' $L$ equal to unity for simplicity (obviously, we can re-scale it as appropriate in our calculations), then
\begin{equation}
 \Psi_\textrm{KS}(x) = \sum_{k=-m}^{m-1}c_k\ \psi_k(x)\ \ \ \ \textrm{with}\ \ \ \ \psi_k(x)=\exp\left(2i\pi k x\right)
 \label{eqn:kSpacePix}
\end{equation}
Since we employ regularly spaced $k$, it follows that the state is periodic: it simply repeats outside of the central box region $-\frac{1}{2}<x<\frac{1}{2}$ but of course, we are only interested in this central region and the confining potential means that the wavefunction amplitudes of the states of interest will vanish well before reaching the border. In fact, we choose a scale such that the wavefunctions have very little amplitude outside of the middle part of this box ($|x|$, $|y|$ less than $1/4$) and this is important because of the form of the Coulomb interaction, as explained in \cref{app: coulomb}.

We employ various basis sizes: for each particle and each dimension, our index runs from $-m$ to $m-1$ for a total number of basis states of $M=2m=2^q$ for some integer $q$. Some prior papers make this choice (e.g. Ref.~\cite{hans_grid_based_2023}) while others have used an odd total number of $k$ states (e.g. Ref.~\cite{RyanLDQSM}) which slightly alters the form of the real-space states.
Such states are found by the Fourier transform of the $k$-space representation; one finds 

\[
\Psi_\text{RS}(x)=\sum_{n=-m}^{m-1}\, b_n \phi_n(x)
\]
where
\begin{equation}
\phi_n(x)=\frac{e^{i \pi (x-p_n)}}{\sqrt M}D_m(2\pi(x-p_n))\ \ \ \text{and} \ \ \ p_n=\frac{n}{M}.
\label{eqn:RSpixel}
\end{equation}
Here $D_m$ is 
\begin{equation}
D_m(u)\equiv\frac{\sin(m u)}{\sin(\frac{u}{2}) }.
\label{eqn:Dexpression}
\end{equation}
One observes that $D_m$ is a variant of the canonical Dirichlet kernel, obtained with substitution $\sin(m)\rightarrow \sin(m+\frac{1}{2})$.  As such, our function is a precursor to the Dirac delta function (or more precisely the Dirac comb  \footnote{A good account of the Dirichlet kernel and its relation to the Dirac comb is provided on https://en.wikipedia.org/wiki/Dirichlet\_kernel.}, since $D_m$ is periodic with period $2\pi$).
Each $\phi_n(x)$ is peaked at $p_n$ and these locations are simply our real-space grid points. Taking our simulation box to run from $x=-\frac{1}{2}$ to $x=+\frac{1}{2}$ then $n$ runs from $-m$ to $m-1$, i.e. $p_n$ runs from $-\frac{1}{2}$ through zero to $\frac{1}{2}-\frac{1}{M}$. Due to the periodicity, the points $\frac{1}{2}$ and $-\frac{1}{2}$ are identical\,\footnote{
The equivalent to \cref{eqn:RSpixel} for Ref.~\cite{RyanLDQSM}, with its odd number of basis states $M$, simply omits the phase prefactor, and has $n$ running from $-m$ to $m$ so that the grid points $p_n$ run from $-(\frac{1}{2}-\frac{1}{2M})$ to $+(\frac{1}{2}-\frac{1}{2M})$.}. Our basis \cref{eqn:RSpixel} is variously referred to as the dual basis, or the real-space basis, or the Dirichlet basis. 

Given that we are initially representing our state in the $k$-space form, \cref{eqn:kSpacePix}, by storing the amplitudes $c_k$ in our simulation software, the transformation to the real-space picture simply requires us to compute the $b_n$ according to 
\[
b_n = \frac{1}{\sqrt{M}}\sum_{k=-m}^{m-1} \exp\left( \tfrac{2i\pi n\,k }{M}\right)c_k.
\]

It is notable that $\phi_n(p_m)=0$ for all $m\neq n$, i.e. where a given basis function is peaked all other basis functions are zero. Consequently it is obvious how to encode any given target wavefunction $f(x)$ into this form; we need only evaluate the target function at the grid points, simply writing
\begin{equation}
f(x)\approx\Psi_\text{RS}(x)=\sum_{n=-m}^{m-1}\,f(p_n) \phi_n(x).
\label{eqn:easyEncode}
\end{equation}

We can naturally extend \cref{eqn:easyEncode} to two or three dimensions with the obvious generalisations
\begin{equation}
    \Psi_\text{RS}(\mathbf{r}) = \sum_{n_x,n_y} f(p_{n_x}, p_{n_y}) \,\phi_{n_x}(x)\,\phi_{n_y}(y)
\label{eqn:2Ddot}
\end{equation}
with $\mathbf{r} = (x,y)$ for 2D dots, and
\begin{equation}
    \Psi_\text{RS}(\mathbf{r}) = \sum_{n_x,n_y,n_z} f(p_{n_x}, p_{n_y}, p_{n_z}) \,\phi_{n_x}(x)\,\phi_{n_y}(y)\,\phi_{n_z}(z),
\label{eqn:3Ddot}
\end{equation}
with $\mathbf{r} = (x,y,z)$ for full 3D. 
Note that we need not have the same basis size $M$ for the various dimensions. Indeed it can be appropriate to have a larger basis along the $x$-axis, which is the axis of the double-dot and therefore has more structure than the $y$-direction. 

Finally we note that the spin degree of freedom is easily included by having one entire charge state of the form \cref{eqn:2Ddot} {\it for each} spin basis state.
 Moreover, for multiple particles, one only needs to take the tensor product of the respective one-particle states. 
 
Although the present paper does not consider the valley degree of freedom, it could be included in the same way as the spin degree of freedom. However, for such a model to be meaningful the $x$-$y$ resolution would also need to be high enough to capture realistic interface roughness since the valley dynamics profoundly depends on such irregularity \cite{cifuentes_bounds_2024}. The computing resources to explore this are beyond the present project, but appear tractable in principle. 

Presently we model a strictly 2D dot using the representation in \cref{eqn:2Ddot}, and we independently compute the full 3D solutions using \cref{eqn:3Ddot}.
The time and memory requirements for numerical computations are of course far larger for the 3D case; for $z$-direction basis size $M_z$ there is a multiplicative cost of order $M_z^2$. Therefore it is interesting to explore the extent to which 2D models can predict the results of a full 3D analysis.  We explore a pseudo-3D approach which requires only the 2D representation. It exploits the fact that we are interested in dots whose $z$-dimension is confinement is tight and relatively uniform, so that we can opt to `freeze out' this axis by assigning only a single basis state. Effectively we assume that the gap to the first excited $z$-state is too large to be relevant to our calculations \cite{burkard_2023, saraiva_2022}. Details of the generalisation are given in Appendix \ref{app: coulomb}. By comparing to the fully fledged 3D solution, we can evaluate the validity of this `short cut'.

Readers may be interested to note that the formalism used here largely follows that in Ref.\,\cite{hans_grid_based_2023}, with some revisions and changes of notation. That paper however focuses on the optimal way to use grid based techniques with future quantum computers; while such machines should avoid the exponential RAM costs, there are downsides including the restriction to unitary evolution which we need not be concerned with. The paper does also include a more complete survey of the history of grid based methods which may be of interest to some researchers.

\subsubsection{Dynamics}

To propagate a state in time we use the split-operator (SO) method. This scheme first splits the total evolution time $t$ in $N$ shorter time intervals $\delta t = t/N$ such that:

\begin{align}
    e^{-iHt} = \underbrace{e^{-iH\delta t}\cdots e^{-iH\delta t}}_{N \; \text{times}}.
\end{align}
Then, as the Hamiltonian in \cref{eq: hamiltonian} can be separated into a kinetic part and a potential part, each short time propagator $e^{-iH\delta t}$ can be approximated using Trotterisation,
\begin{align}
    e^{-iH\delta t} &= e^{-i(H_{kin}+H_{pot})\delta t} \nonumber \\ 
    &= e^{-iH_{pot}\delta t/2}e^{-iH_{kin}\delta t}e^{-iH_{pot}\delta t/2} + O(\delta t^3).
\end{align}
We find that second-order Trotter is sufficient for our purposes. 

To obtain the time-evolved state $\ket{\psi(t)}$, we need to apply the operator $U_{SO}(\delta t) := e^{-iH\delta t}$ to the initial state $\ket{\psi(0)}$ $N$ times. In the grid-based formalism this can be done efficiently by switching between the \emph{real-space} and \emph{k-space} representations. Indeed, due to our choice of encoding, $H_{kin}$ is diagonal in \emph{k-space} while $H_{pot}$ can be approximated as diagonal in \emph{real-space} (see \cref{app: coulomb}).

\section{\label{sec: exchange} Exchange coupling}
In this section we consider the practical task of estimating the exchange coupling $J$ of a DQD structure. As $J$ is directly related to the speed of two-qubit gates, any device that cannot generate a substantial exchange coupling is unsuitable as a qubit-qubit gating system.

The $J$ coupling causes the spin singlet state to acquire phase with respect to all triplet states, and thus induces a continuous \text{SWAP} operation for our two qubits. The effect is not due to any direct (e.g. dipolar) interaction between spins, but rather it is a consequence of fermionic particle exchange antisymmetry: the spin singlet state must be associated with a symmetric charge state, and conversely spin triplet states with antisymmetric charge states. These charge states are appreciably distinct in energy, and the energy gap is highly sensitive to the size and shape of the confining double dot potential. 

To be specific, for two electrons trapped in a DQD the exchange coupling is defined as the difference in energies between the first two eigenstates of the spinless two-electron system (which indeed have opposite exchange symmetry). These eigenstates can be obtained by diagonalising the Hamiltonian within our real-space basis. However, in the case of two electrons in 2D with $M$ grid points per particle per direction (a total of $M^4$ basis states), the diagonalisation becomes rapidly intractable. Nevertheless, by exploiting the distinct symmetries of the eigenstates with respect to electron swapping and employing a state-preparation technique known as imaginary-time evolution (see e.g. \cite{lethovaara_2007}), the grid-based method provides an efficient way of preparing them.

\begin{figure*}
    \centering
    \includegraphics[width=1.\textwidth]{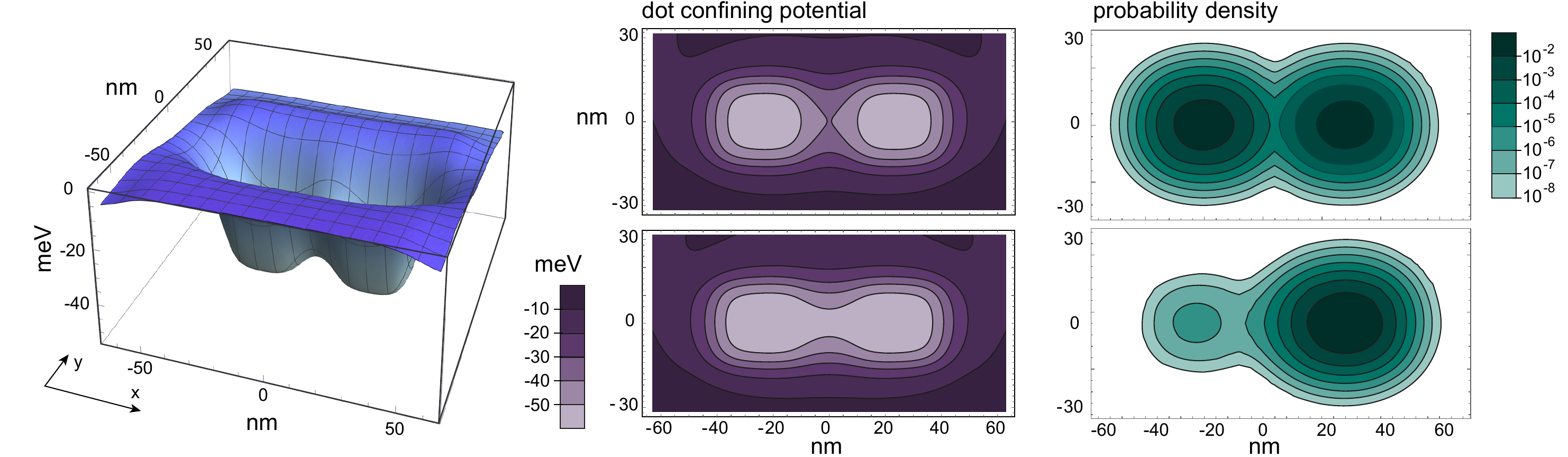}
    \caption{ 
    Plots showing the potential well constituting the double dot (left and centre) and states of the confined electrons (right). The 3D visualisation and the upper contour plot show the `barrier up' case, while the lower contour plot shows the `barrier down' potential. The right-most plots show probability densities on a logarithmic scale. The upper-right plot shows the ground state probability density for the `barrier up' scenario (due to symmetry the distribution is identical for either particle). We then explore the subspace corresponding to ``a given electron is definitely on the left of the structure'' and plot the probability distribution for {\em the other electron}; this reveals a small but non-zero probability that the second particle is also on the left, i.e. a (2,0) charge state which would not constitute a legitimate qubit configuration. The total probability of (2,0) and (0,2) is 0.04\%.}
    \label{fig:state_prep}
\end{figure*}

\subsection{State preparation}

Intuitively, for two electrons in a double quantum dot, the ground and first excited states should be close to a symmetric and antisymmetric combinations of finding one electron on the left and the other on the right respectively. 

Our approach to initializing the system into its ground state and first excited state (the lowest lying symmetric and antisymmetric charge states, respectively) is simply an imaginary time evolution. This is convenient as it exploits the same machinery that we use for modelling the dynamics of the system; we need only adjust a line in our code to alter from $i\, t\rightarrow t$ in the Schrödinger equation, together with a convergence test and an adaptive $\delta\,t$.

Starting from a guess initial state which has some overlap with the true ground state -- such as a symmetric combination of two Gaussians centered in each dot -- propagating the state using an imaginary time guarantees convergence to the true ground state $\ket{s}$ \cite{lethovaara_2007}. The ground and first excited states have distinct symmetries and imaginary-time evolution preserves the symmetry of the starting state. Hence, starting with an antisymmetric combination of Gaussians, we will converge to the antisymmetric state with the lowest energy namely the first excited state $\ket{a}$. We can then estimate $\ket{s}$ and $\ket{a}$ respective energies by computing their expectation value with respect to the Hamiltonian and then determine the exchange coupling by computing their difference. This method is completely general and can be applied to any DQD structure.

For realistic potentials, we will find that the $\ket{s}$ and $\ket{a}$ include a finite probability of finding both electrons in the same dot. However, when the barrier is large, we expect to be able to approximate them by,
\begin{align}
    \ket{s} &\approx \frac{1}{\sqrt{2}}\left( \ket{\text{LR}} + \ket{\text{RL}}\right), \nonumber \\
    \ket{a} &\approx \frac{1}{\sqrt{2}}\left( \ket{\text{LR}} - \ket{\text{RL}}\right),
    \label{eq: symm_antisymm_approx}
\end{align}
with the probability of finding the electrons in the states $\ket{\text{LL}}$ and $\ket{\text{RR}}$ decreasing exponentially with the barrier height. Although this probability should become small, it is important to characterise it since it represents states for which qubits are not defined -- finding the system in such a state would correspond to irreversible qubit loss. 

Throughout the paper we will use the state $\ket{\text{L}}$ ($\ket{\text{R}}$) to denote the state of an electron that is definitely on the left (right) dot. This notation will especially be useful when defining our qubit states in \cref{eq:qubit_states}. However, it is important to note that when approximating the symmetric and antisymmetric state as in \cref{eq: symm_antisymm_approx} we are referring to specific charge wavefunctions (the eigenstates) localised on each dot. 

\subsection{Numerical results}
\begin{figure}
    \centering
    \includegraphics[width=\linewidth]{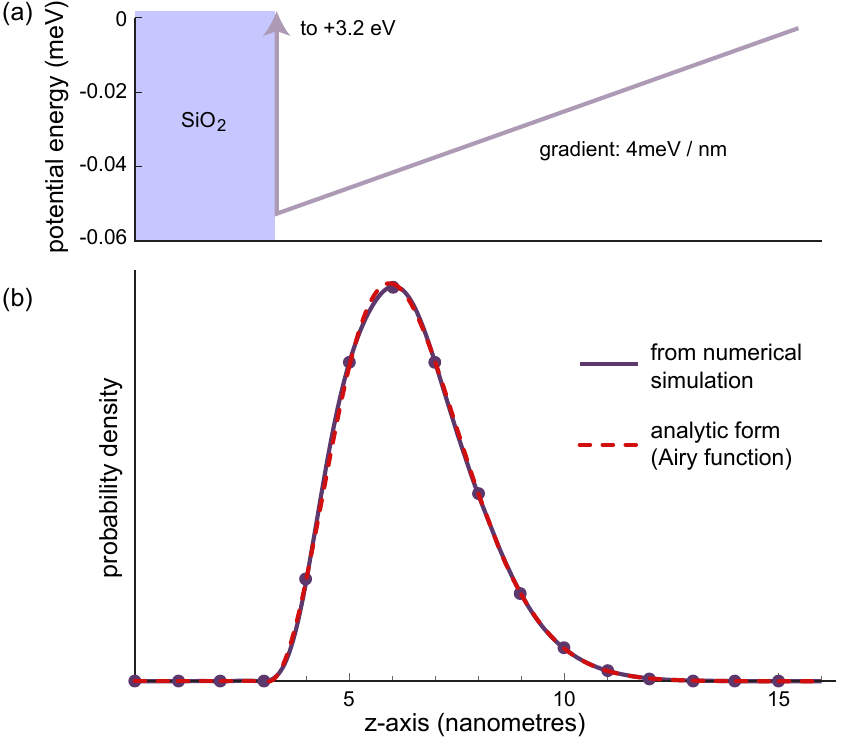}
    \caption{Panel (a) shows the variation in potential within the 3D double dot in the $z$-direction (the $x$,$y$ location corresponds to a dot minimum). The potential is linear, ramping at 4 meV per nanometre. Panel (b) shows the probability density for an analytic Airy function solution to a 1D problem with potential from (a), but taking the step to be infinite. One sees that this analytic form is a near-perfect match to the probability density that is obtained from our {\it ab initio} calculation to the full 3D quantum dot potential.}
    \label{fig:Airy_matching}
\end{figure}

We computed $J$ using the {\it ab initio} method described in the previous section, for different basis sizes, choices of dot dimensionality, and representations of the Coulomb interaction. For each such choice, we swept through a range of potentials from `barrier up' to `barrier down' (see Fig.~\ref{fig:state_prep}); the potentials were obtained from a commercial Poisson-solver tool provided with a certain device layout involving a central electrode (see \cref{fig:main_figure} and \cref{app:device_spec}). Throughout our numerical study, where we use a 3D potential this is created by introducing a $z$-direction ramp (see Fig.~\ref{fig:Airy_matching}(a)) to the corresponding 2D potential. This choice was made in order that the behaviour of the 2D and 3D models can be compared directly without factors arising from, e.g, small variations in the $z$-direction field across the $x$-$y$ plane that would occur in reality. The 3D numerical method does not depend on this simple structure, so that its runtime and accuracy should be similar for any smoothly varying 3D potential. 

\begin{figure}
    \centering
    \includegraphics[width=\linewidth]{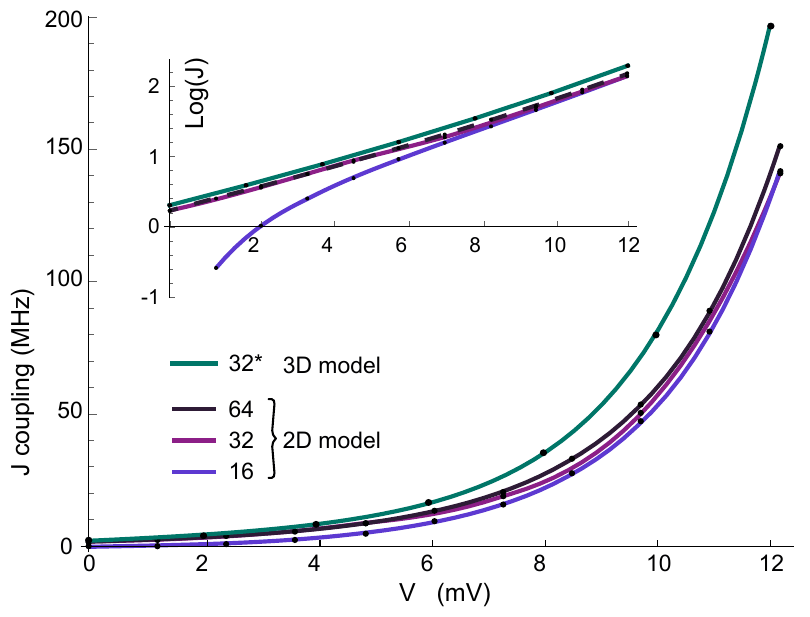}
    \caption{Observed variation of the exchange coupling $J$ with the barrier gate voltage $V_g$. The blue and purples lines correspond to a strictly 2D dot modelled with 16, 32 and 64 basis states per particle, per dimension. One notes from the inset that curves for 32 and 64 are near-identical (the 64 line is shown dashed for visibility), while the lower basis size of 16 deviates in the low $J$ limit. The green curve corresponds to the far more computationally demanding 3D model; the basis size is 32 in the $x$-direction (the axis along which the two dots lie) and 16 in the orthogonal directions. One sees that the 2D model underestimates the $J$ coupling strength appreciably.}
    \label{fig:exchange_couplingA}
\end{figure}

Our first set of results are shown in  \cref{fig:exchange_couplingA}. This figure compares several 2D dot calculations (with various basis sizes) with the far more computationally demanding 3D calcualtion. In all cases we retrieve the general exponential dependence of the exchange coupling with the barrier gate voltage, as expected from analytic treatments (see e.g. \cite{burkard_2023}). The purple lines depict cases for a strictly 2D dot; while noting broadly similar behavior for all resolutions, we observe that for the $16^4$ case, there is a deviation from the expected behaviour at low $V_g$ (the `barrier up' scenario). The use of $32^4$ is adequate to recover the expected exponential variation over the full range of $V_g$ considered. In fact further calculations (not shown) verified that a resolution of $(32\!\times\! 16)^2$ suffices to obtain good low-$J$ estimates, provided that larger basis runs along the $x$-axis (see Fig.\,\ref{fig:state_prep}, leftmost panel).

The green line is the data from the computationally-expensive full 3D dot simulation. The potential differs from the 2D case by the addition of a $z$-direction component that ramps linearly up with depth, and an abrupt interface `wall', see Fig.\,\ref{fig:Airy_matching}(a). The grid-based model used a total basis size of $(32\times\!16\!\times\!16)^2\sim 67\times10^6$, where the larger basis size $32$ runs long the $x$-axis (the axis along which the two minima of the double dot lie). By repeating the 3D calculation for somewhat higher resolutions we confirmed that there was negligible variation in the predicted $J$. 

Intuitively, it is not unexpected that the introduction of finite width is significant. Recall that the Coulomb interaction actually {\it reduces} the exchange coupling, by forcing electrons further apart and reducing their amplitude in the crucial central region where the state's symmetry is most significant. By allowing the electrons to move in a third dimension we reduce the impact of the Coulomb interaction and thus increase the exchange coupling \cite{burkard_2023}. 

We investigated the extent to which a modified 2D model could replicate the $J$ strength predictions of the full 3D model. We began by inspecting the probability density in the $z$-direction, according to the full wavefunction solution. Specifically, we compared the numerically-obtained probability density with the analytic solution to a 1D model in which the interface step is infinite (noting that the finite step used in the 3D model, 3.2 eV, is extremely severe). The numerically obtained solution is remarkably close to the analytic form, see Fig.\,\ref{fig:Airy_matching}(b). We therefore proceeded to adjust the 2D Coulomb interaction using the analytic Airy function solution, noting that process does not use any information gained from the full 3D model. The method we employed is explained in \cref{app: coulomb}. In essence, we derive a whole set of 2D Coulomb interaction matrices, each of which is has a fixed $\Delta z$ offset between the two particles. Then we find the average of these matrices, weighted by the Airy-function probability distribution. This has the effect of softening the Coulomb interaction at short range. 

\begin{figure}
    \centering
    \includegraphics[width=\linewidth]{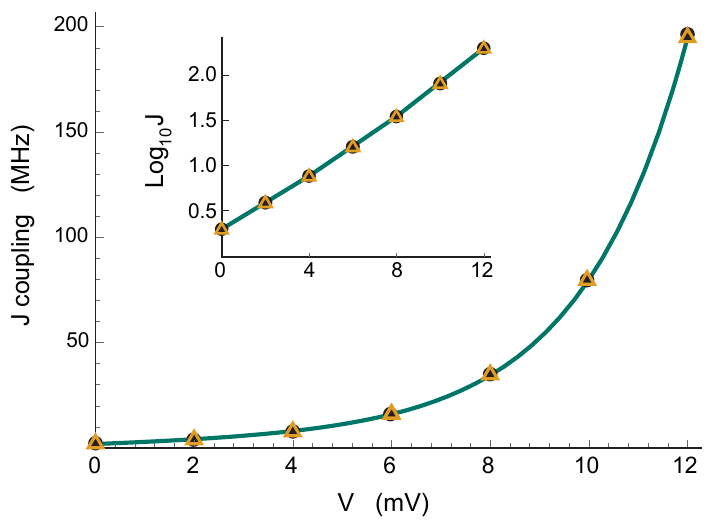}
    \caption{Performance of our resource-efficient quasi-3D model, which uses a 2D space and an adjusted Coulomb interaction. The black dots correspond to the true 3D model, and are the same data as in the green line of Fig.\,\ref{fig:exchange_couplingA}. The triangles denote the predictions made from the quasi-3D model. One sees that the match is near-perfect; in fact the quasi-3D model slightly underestimates $J$ by deviation of between $0.6\%$ and $0.8\%$.}
    \label{fig:augmented2D_versus3D}
\end{figure}

Figure\,\ref{fig:augmented2D_versus3D} compares the $J$ versus $V_g$ curves for 2D versus 3D dots. One observes that the match is excellent, indicating that for dots of this kind -- i.e. with negligible irregularity in the $z$-confinement over the lateral extend of the dot -- this approach of quasi-3D can offer an excellent trade off of accuracy versus numerical efficiency. 

\color{black}

Because the assignment of a mean-$z$ depth of the electrons is a choice that makes our models specific to only dots of that geometry, in the rest of the paper, we will use the pure 2D Coulomb interaction. While noting of course that this is in some sense the least `realistic' option, we emphasise that the phenomena which we presently identify are qualitatively robust to the introduction of a small finite $z$-depth: There is no qualitative impact on the physics observed during the dynamics. The only meaningful change will be the speed of the gates. Moreover, it is straightforward to recalculate any given set of results for a specific $\langle z \rangle$.

\section{\label{sec: time_dynamics} Exchange gate dynamics}
We now focus on the dynamics of our two-electron system. Specifically, we are interested in extracting two-qubit gate fidelities and loss probabilities for different voltage profiles. But first, we need to define our qubits' states.

\subsection{Qubit states}
For two electrons in a DQD in the $(1,1)$ charge configuration, the qubit states are defined as follows, 
\begin{align}
    \ket{00} &\coloneqq \frac{1}{\sqrt{2}}\left( \ket{\text{LR}} - \ket{\text{RL}}\right)\ket{\downarrow\downarrow}, \nonumber \\
    \ket{01} &\coloneqq \frac{1}{\sqrt{2}}\left( \ket{\text{LR}}\ket{\downarrow\uparrow}  -\ket{\text{RL}}\ket{\uparrow\downarrow}\right), \nonumber \\
    \ket{10} &\coloneqq \frac{1}{\sqrt{2}}\left( \ket{\text{LR}}\ket{\uparrow\downarrow} - \ket{\text{RL}}\ket{\downarrow\uparrow}\right), \nonumber \\
    \ket{11} &\coloneqq \frac{1}{\sqrt{2}}\left( \ket{\text{LR}} - \ket{\text{RL}}\right)\ket{\uparrow\uparrow}.
    \label{eq:qubit_states}
\end{align}
where the arrows represent the spin states of the electrons.
The state of a qubit is given by the spin of an electron found in a specific dot. For instance, $\ket{01}$ represents the state in which we have an electron with spin down on the left dot and an electron with spin up on the right dot. This state is naturally given by the antisymmetrisation of the distinguishable-electrons state $\ket{\text{LR}}\ket{\downarrow\uparrow}$ in which we find the first electron on the left with spin down and the second electron in the right dot with spin up.
Note that the singlet $S(1,1)$ and triplet $T_0(1,1)$ states can be retrieved as follows,
\begin{align}
    S(1,1) &= \frac{1}{\sqrt{2}}\left(\ket{01} - \ket{10}\right), \nonumber \\
    T_0(1,1) &= \frac{1}{\sqrt{2}}\left(\ket{01} + \ket{10}\right).
\end{align}

In order to study two-qubit gates within our model, we need to be able to initialise our electrons in the qubit states defined in \cref{eq:qubit_states}. However it is important to appreciate that the valid qubit states defined in \cref{eq:qubit_states} will not suffice to completely describe the state of a real double-dot structure. This is because the true eigenstates of the system, $\ket{s}$ and $\ket{a}$, are only imperfectly described using the $\ket{\text{LR}}$, $\ket{\text{RL}}$ states (with the imperfect association as given in \cref{eq: symm_antisymm_approx}). In reality there is always some finite probability associated with $(2,0)$ and $(0,2)$ charge configurations, i.e.  $\ket{\text{LL}}$, $\ket{\text{RR}}$, which lie outside the qubit subspace and thus constitute qubit loss outcomes. With that said, the approximate association is 
\begin{align}
  \ket{00} &\approx \ket{a}\ket{\downarrow\downarrow}, \nonumber \\
  \ket{01} &\approx \frac{1}{2}\left( \left(\ket{s} + \ket{a}\right)\ket{\downarrow\uparrow} - \left(\ket{s} - \ket{a}\right)\ket{\uparrow\downarrow} \right), \nonumber \\
  \ket{10} &\approx \frac{1}{2}\left( \left(\ket{s} + \ket{a}\right)\ket{\uparrow\downarrow} - \left(\ket{s} - \ket{a}\right)\ket{\downarrow\uparrow} \right), \nonumber \\
  \ket{11} &\approx \ket{a}\ket{\uparrow\uparrow},
\label{eq: init_qubit_states}
\end{align}
with the approximation becoming better as the barrier gets larger. This encoding naturally enforces a coherent oscillation (with frequency $J$) between the states $\ket{01}$ and $\ket{10}$. 

As noted earlier, adding the spin degree of freedom is done by taking the tensor product of the charge states defined in \cref{eqn:2Ddot} with an array of length two encoding the spin state. The generalisation to two particles is straightforward. It is worth noting that as we are using a first-quantised representation, the antisymettry of the wavefunction is enforced in the states and is conserved through the evolution in time.

\subsection{Numerical results}
While there are multiple two-qubit gates made possible from the exchange interaction \cite{meunier_2011}, in this work we focus on the entangling $\sqrtswap$ gate {which naturally appears when considering a homogeneous magnetic field. Initially we will thus suppose that the $g$-factor in each dot is the same. In that case, the gate is realised by pulsing the barrier gate voltage such that $\int_0^{T_{\sqrtswap}} J(t) dt = \frac{\pi}{2}$.}
In \cref{sec: robust_pulse} we will study the impact of having different $g$-factors. For simplicity, we focus on simple pulse sequences in which we smoothly ramp the potential from one with a large barrier (exchange off) to one with a small barrier (exchange on). It is worth noting that our tool can handle any evolution of the potential with time.

\begin{figure}
    \centering
    \includegraphics[width=\linewidth]{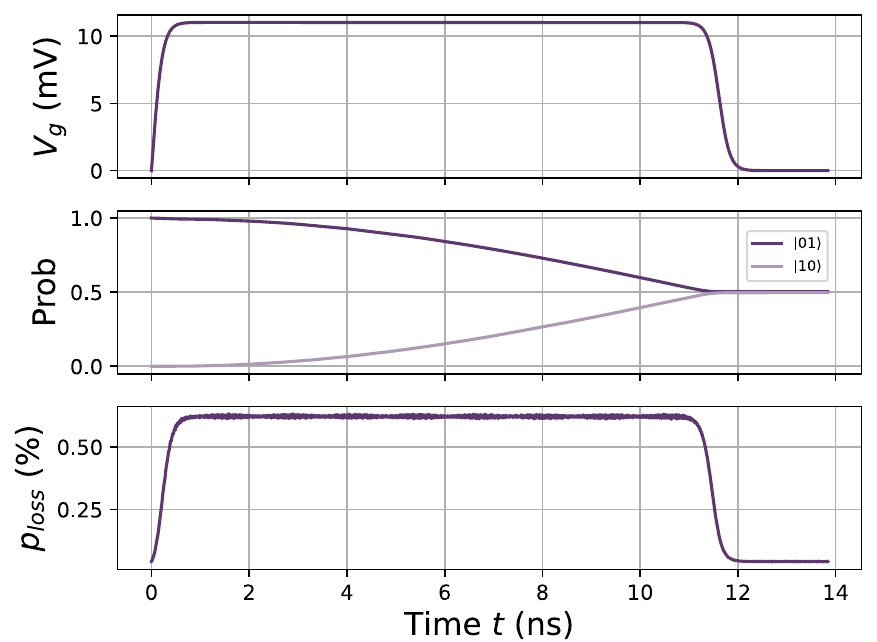}
    \caption{Smooth $\sqrtswap$ pulse. (top) Barrier gate voltage profile with $V_g$ being the difference in potential between the bottom of the wells and the barrier at the midpoint $(x=y=0)$. We smoothly switch on the exchange interaction to minimise loss. (middle) Probability to find the electrons in the $\ket{01}$ (dark) and $\ket{10}$ (light) states. We observe a near perfect transfer to the desired superposition at the end of the process with an infidelity $1-F = 1.32 \times 10^{-5}$. (bottom) Probability $p_{loss}$ to find the electrons in the charge $(2, 0)$ or $(0, 2)$ configurations and hence outside of the qubit subspace. It is worth noting that for our initial state, we have a small probability $p_{loss} = 0.04 \%$ to be in the $(2, 0)$ or $(0, 2)$ configurations. At the end of the process, we recover a practically identical value for $p_{loss}$ and conclude that our adiabatic pulse doesn't increase the loss (even though if one were to perform a measurement of the charge distribution {\it midway} through the gate it {\it would} cause appreciable loss probability).}
    \label{fig:smooth_swap}
\end{figure}

Starting from the state $\ket{01}$ defined as in \cref{eq: init_qubit_states}, we perform a $\sqrtswap$ operation by dynamically changing the barrier gate voltage according to the schedule given \cref{fig:smooth_swap}. The rising and falling times were chosen to prevent non-adiabatic effects (as discussed later) while minimising gate time -- ramping over the course of approximately one ns is consistent with current technological capabilities.

After the dynamics are simulated, we must measure the performance of the gate. Finding that the state at the end of the gate satisfies $\text{Prob}(\ket{01}) = \text{Prob}(\ket{10}) \approx \frac{1}{2}$ is of course insufficient to conclude that we indeed perform a $\sqrtswap$ gate. To characterise the quality of the gate we need to introduce more nuanced measures. Suppose we are given a general quantum state $\ket{\psi}$ encoded in real-space as follows:
\[
\ket{\psi} =\sum_i \ket{\Omega_i}(
a_{i}^{\downarrow\downarrow}\ket{\downarrow\downarrow} + a_{i}^{\downarrow\uparrow}\ket{\downarrow\uparrow} + a_{i}^{\uparrow\downarrow}\ket{\uparrow\downarrow} + a_{i}^{\uparrow\uparrow}\ket{\uparrow\uparrow}
)
\]
where the constants can have any values provided that overall antisymmetry is respected. 
Here the set of states $\{\ket{\Omega_i}\}$ is some relevant basis of two-particle charge states, for example the grid-based Dirichlet basis used in our model.

We wish to measure how close this state is to a given qubit state $\ket{Q}$ which would typically be written in the form,
\begin{align}
    \ket{Q} &= a_{00}\ket{00} + a_{01}\ket{01} + a_{10}\ket{10} + a_{11}\ket{11} \nonumber \\
    =&\frac{1}{\sqrt 2}\big( \ket{\text{LR}}\left( a_{00}\ket{\downarrow\downarrow} + a_{01}\ket{\downarrow\uparrow} + a_{10}\ket{\uparrow\downarrow} + a_{11}\ket{\uparrow\uparrow}\right) \nonumber \\
    &- \ket{\text{RL}}\left( a_{00}\ket{\downarrow\downarrow} + a_{01}\ket{\uparrow\downarrow} + a_{10}\ket{\downarrow\uparrow} + a_{11}\ket{\uparrow\uparrow}\right)\big).
    \label{eq:TargetQubit}
\end{align}
(Notice the different kets associated with $a_{01}$ in each line, and similarly for $a_{10}$).
In our target state we have used the conventional $\ket{\text{LR}}$ notation to indicate that electron 1 is definitely in the left side of the structure, and electron 2 is definitely in the right (and conversely for $\ket{\text{RL}}$). States such as $\ket{\text{LL}}$ or $\ket{\text{RR}}$, with both electrons on the same side, are not legitimate qubit states.
Therefore in measuring the match between $\ket{\psi}$ and $\ket{\phi}$ we can start by projecting out any $\ket{\Omega_i}$ which correspond to the $(2,0)$ and $(0,2)$ configurations, recording the probability to be in this non-meaningful subspace as $p_\textrm{loss}$. 

The fidelity {\em within} the qubit subspace is then to be determined. Importantly, we are not concerned with the detailed shape of the charge distribution, thus for example if the $a_i$ coefficients in \cref{eq:TargetQubit} are correct then the fidelity is unity regardless of the specifics of the charge state $\ket{\text{LR}}$ (and its label-exchanged converse $\ket{\text{RL}}$). Therefore we write   
\[
F(\ket{Q}, \ket{\psi}) \coloneqq \sum_{\textrm{all}\,\ket{\text{LR}}}|\braket{Q|\psi}|^2
\]
where we are summing over all variants of the target $\ket{Q}$ defined by \cref{eq:TargetQubit} using a complete basis of states $\ket{\text{LR}}$ (i.e. all the states with a $(1,1)$ charge distribution). Fortunately the calculation of this fidelity in the grid-based representation is straightforward. 

This expression gives us the fidelity of a given state. The fidelity of the gate itself follows relatively straightforwardly simply by inspecting the state fidelity (versus the ideal) of two cases: that of an input state $\ket{Q_{\text in}}$ that is symmetric under qubit exchange, and a $\ket{Q_{\text in}}$ that is antisymmetric. This suffices to infer the state fidelity of any output, and hence the gate fidelity under any chosen definition. We chose the definition of gate fidelity as being the worst case fidelity, i.e. the state fidelity for input $\ket{Q_{\text in}}$ which minimises $F(\langle Q_{\text out}, \ket{\psi})$ where $\ket{Q_{\text out}}$ is the ideal output. This will be any equal superposition of a symmetric and an antisymmetric qubit state, and thus the state $\ket{01}$ itself is suitable, see \cref{eq: init_qubit_states}.

In this case, our pulse should ideally achieve the target state $\ket{Q} = \tfrac{\ket{01} - i\ket{10}}{\sqrt{2}}$. As noted earlier, if we begin from a state in the lowest two eigenstates then there is a small initial amplitude associated with the states having a $(2,0)$ or $(0,2)$ charge distribution. For the optimised gate, the eventual $p_\textrm{loss}$ corresponds only to this initial imperfection. The increase of $p_{\text{loss}}$ during the gate is due to the fact that in this exchange coupling regime, the overlap with states outside the computational subspace is bigger (which is inevitable to obtain large $J$). The action of the gate itself introduces {\em zero loss probability} (to within the numerical noise in the model $\sim10^{-7}$). Moreover the infidelity within the qubit subspace process is very small, $1-F = 1.32 \times 10^{-5}$.  We thus verify that with the chosen voltage profile, one achieves a near-perfect $\sqrtswap$ gate when starting from $\ket{01}$ (and by symmetry, also when starting from $\ket{10}$). 
As a comparison, if we abruptly switch the barrier (see \cref{app:abrupt_transition}), $p_{loss}$ is dramatically increased and can exceed $1\%$, and even once this component is removed we obtain only the limited fidelity $F = 97.9 \%$. This underscores the importance of a proper voltage switching profile, i.e. achieving near-adiabatic dynamics.

Whereas the ideal system can be adjusted to very high (near-perfect) gate operation, this does not account for imperfect environments -- we now discuss the impact of charge irregularities. 

\section{\label{sec: charge_noise} Impact of charge noise on exchange gate}
To adequately model the noise hindering the performance of silicon spin qubits, we first need to understand its nature. Multiple experimental groups have performed noise spectroscopy and they found that it follows a $1/f$-like behaviour \cite{connors_2022, yoneda_2018, wuetz_2023, rojas-arias_2023}. Such noise is commonly associated with charge noise, and it is believed that it can be well-modelled using a large number of two-level fluctuators \cite{elsayed_2022}. Several works have considered baths of fluctuators \cite{elsayed_2022, kepa_2023, mickelsen_2023} to model this regime. Recent works \cite{rojas-arias_infering_2025} have shown that, for some devices, only a few two-level fluctuators are present, breaking the $1/f$-like model. We complement prior modelling works by exploring the interesting case of a single fluctuator strongly coupled to the DQD \cite{cifuentes_2023, shehata_2022}. We leave more diverse TLFs (eg. dipoles) for future work.

As shown in \cite{shehata_2022}, considering the time it takes to perform a single $\sqrtswap$ gate, one expects that there will be more static charge defects than fluctuating charges within the period of a single gate operation. Hence we will first limit ourselves to static charges, before looking at the more rare, albeit interesting, case of fluctuating charges. 

\subsection{Static charge}
The trapped charge is modelled classically by adding an extra coulomb interaction to \cref{eq: hamiltonian} with respect to each electron in the DQD.
\begin{align}
    H_{sc, i}(\mathbf{r})= \frac{e^2}{4\pi\epsilon_0\epsilon_r|\mathbf{r}-\mathbf{r_i}|^2},
\end{align}
with $\mathbf{r} = (x, y, z)$ and $\mathbf{r_i}$ the position of the charge and the position of the $i^{th}$ electron respectively. The charge is randomly placed at the silicon oxide interface $z \approx 2$ nm above the DQD and in the symmetry axis of the double dot (the $x$ axis, i.e. $y = 0$). By placing the charge along this axis, we are exploring situations in which the charge has the most impact. In \cref{app:y_dir_charge}, we explore the more general setting of off-axis charge defects. 

When a single charge is located near a DQD, it displaces the trapped electrons, altering the overlap between their wavefunctions and hence modifying the exchange coupling. This increases as the charge gets closer to the DQD (see \cref{tab:exchange_noise}). However, the trend breaks down when the charge is between the two electrons. We assume that the gate has been calibrated in the absence of a charge defect, i.e. as operators we are unaware of it and thus we employ the same pulse as in the noiseless case to perform our gate. As the exchange coupling is modified by the presence of the charge, the pulse no longer yields a high-quality gate.

\begin{figure}
    \centering
    \includegraphics[width = \linewidth]{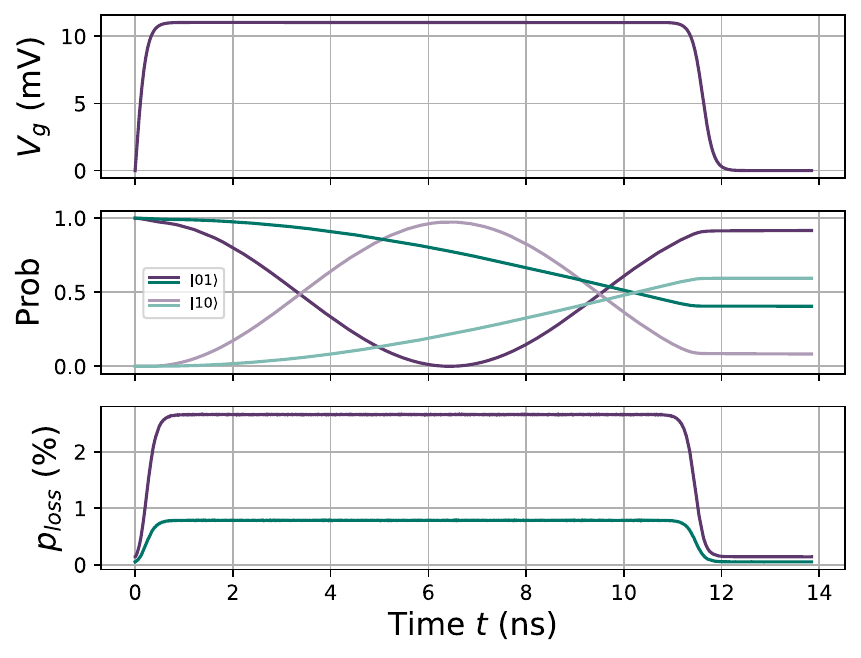}
    \caption{Smooth $\sqrtswap$ pulse in the presence of charge noise. (top) As we assume the operator does not know that the defect charge is present, we use the evolution of the barrier gate that works optimally in the defect-free case (i.e. upper the panel of this figure is the same as in Fig.~\ref{fig:smooth_swap}). (middle) Probability to find the electrons in the $\ket{01}$ (dark) and $\ket{10}$ (light) for a charge positioned at $x = 50$ (purple) and $x = 100$ (green) nm respectively. The presence of charge noise generates an error in the $\sqrtswap$ gate as it modifies $J$. There is also an increased value for $p_{loss}$ during the gate operation, as compared to the noiseless gate (Fig.\,\ref{fig:smooth_swap}), but at the end of the gate it returns to the initial value and thus the operation does not {\it increase} the loss -- it is still effectively adiabatic.}
    \label{fig:noisy_swap}
\end{figure}

In \cref{fig:noisy_swap}, we report the evolution of the qubits' state and $p_{loss}$ for a charge positioned at $x = 50$ nm (purple) and $x = 100$ nm (green) with respect to the center of the double dot structure ($x=y=0$). Assuming that the system has previously relaxed into the eigenstates of the modified potentials, our initial state $\ket{01}$ differs from that computed earlier and now has a larger initial value of $p_{loss}$. We note that the control pulse still gives us an adiabatic evolution. Nevertheless, we find that the fidelity decreases drastically with charge proximity because of the timing error. In \cref{tab:exchange_noise}, we report the evolution of $F$ and $p_{loss}$ for various charge positions. 

\begin{table}[]
    \centering
    \begin{tabular}{c|c c c c c}
        \hline
        \hline
        Charge position (nm) & 50 & 75 & 100 & 150 & 200 \\
        \hline
       $J$ (MHz) & 507 & 202 & 168 & 155 &  153 \\
       $p_{loss} \; (\%)$ & 0.14 & 0.06 & 0.05 & 0.05 &  0.04\\
       $F \; (\%)$&  22.41 & 92.11& 99.10 &  99.95&  99.99\\
       \hline
       \hline
    \end{tabular}
    \caption{Exchange coupling, loss probability and state-to-state fidelity as a function of the position of the charge. The charge position (in the dot line) is given as a distance from the center of the DQD which is positioned at (0, 0, 0). As the exchange coupling is a static property, it has been computed with the $32^4$ resolution. However, as the loss and fidelities are obtained after simulating the dynamics, we only used a $16^4$ resolution which we don't expect to change much (see \cref{fig:exchange_couplingA})}. 
    \label{tab:exchange_noise}
\end{table}

\subsection{\label{subsec: fluc_charge} Fluctuating charge}
Although the majority of charge defects may be static on the timescale of a $\sqrtswap$ operation, it is important to study the impact of a charge that moves during a gate operation. While there are a vast range of interesting scenarios that can studied by the methods we use here (e.g. a charge moving randomly in 3D), we will focus on a charge fluctuating instantaneously between two traps locations. 

We take two approaches to exploring the effect. In the first, we specify that the charge oscillates between $\mathbf{r_{init}} = (100, 0, 2)$ nm (its initial position) and $\mathbf{r} = (x, 0, 2)$ nm for various values of $x$. 
We expect that the impact will be greater as the two positions are far from one another. We will also suppose that the experiment has been calibrated such that one has modified the barrier gate voltage profile such that we obtain an optimal $\sqrtswap$ gate in the presence of a charge at $\mathbf{r_{init}}$. In the second approach, described presently in the context of robust pulses, we consider fluctuators with a fixed 10\,nm separation but various mean locations. A charge could tunnel between nearby defects on the scale of a few nm. Here, we consider slightly worse case to show the power of our robust pulse sequence.

The nature of the TLFs that will be most problematic for quantum processes is an active topic in the literature, and the mismatch between whole-charge fluctuations and experimental power spectral densities have been documented in \cite{shehata_2022}. Here our aim it to demonstrate the flexibility of our numerical tool by investigating one potentially-damaging form of TLF.

\begin{figure}
    \centering
    \includegraphics[width = \linewidth]{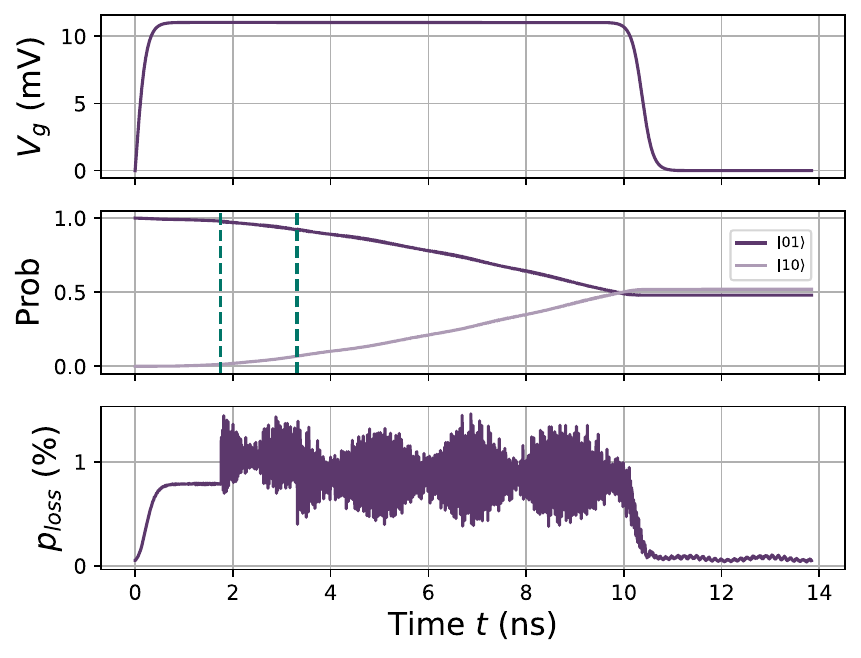}
    \caption{Smooth $\sqrtswap$ pulse in the presence of a fluctuating charge. The charge oscillates between its initial position $\mathbf{r_{init}} = (100, 0, 2)$ nm and $\mathbf{r} = (75, 0, 2)$ nm at each green dashed lines.}
    \label{fig:fluc_example}
\end{figure}

In \cref{fig:fluc_example}, we study the impact on the gate of a charge oscillating between $\mathbf{r_{init}}$ and $\mathbf{r} = (75, 0, 2)$ nm. We randomly chose the times (dashed green lines) at which the charge moves and we limited the number of fluctuations to at most three (see \cref{app:charge_fluc} for other examples). We limit ourselves to three as these scenarios become less and less probable as the number of fluctuations increases.

It is interesting to note that instantaneously changing the position of the charge during the gate has a similar effect as when we too-abruptly switch the barrier voltage \cref{app:abrupt_transition}. The sudden (non-adiabatic) nature of the change is damaging. However, as the charge doesn't move too closely to the DQD in this example, the change in potential is small here, and hence the impact of the fluctuation is relatively minor (see the last row of \cref{app:charge_fluc} for more drastic effects). Regarding $p_{loss}$ we note that even after the end of the gate, we still have some oscillations. We take the average over the time remaining after the gate and find that $p_{loss} = 0.07 \%$ and $F = 99.71 \%$. Contrary to the previous cases, we now for the first time seen an increase to $p_{loss}$, which is the worst type of error as it cannot be corrected. A fluctuating charge is thus more damaging than a static one when it comes to gate quality.

Now that we have identified how charge noise modifies the quality of the gate we access the vast literature on robust pulse sequences to begin to explore methods to mitigate these errors.

\section{\label{sec: robust_pulse} Robust pulse sequences}
\subsection{Types of errors}
When performing a gate, three types of imperfection can arise. We already saw two of them in the previous section. First, the loss coming from driving the electrons into the $(0,2)$ and $(2,0)$ configurations (and thus outside of the qubit subspace) when modifying the potential. Second, the error generated by charge noise which modifies the exchange coupling $J$ between electrons, and ultimately leads to uncertainty in the gate time, lowering the fidelity of the operation. Finally, a non-ideality that we neglected until now arises from magnetic detuning between dots. If the $g$-factors between the dots are different, the qubits within the double dot will not experience a pure SWAP-type operation: Consider a Bloch sphere for the qubit subspace $\ket{\text{01}}$, $\ket{\text{10}}$ (as defined in \cref{eq:qubit_states}, then rather than an on-axis rotation, in the presence of a $g$-factor discrepancy the qubit state will rotate around a tilted axis. We now consider this effect i.e. we assume each dot's $g$-factor is indeed different. For simplicity we assume that the $g$-factor is uniform across each half of the double dot structure, changing abruptly at the center (i.e. at $y=0$). This is obviously oversimplified, but allows us to make contact with prior treatments -- note however that the numerical methods described here can tackle any spatially varying $g$-factor.

While in the case of a well-characterised environment, the first type of error (qubit loss) can be reduced by using an adiabatic pulse, we observed that in the presence of unknown charge fluctuations, it cannot be corrected. However, as we will explore below, the other types of errors can be mitigated even in the presence of unknown charge environments.

It is also worth observing that a two-qubit gate that is off-axis in the sense described above is nevertheless adequate as a computationally universal gate, given suitable single-qubit rotations. Therefore one could argue that the natural rotation is acceptable (once characterised) and does not require correction at the pulse level. However, in a device with thousands or millions of such structures the ability to homogenize their action is likely to be valuable. 

\begin{figure*}
\centering
\includegraphics[width = 0.98\linewidth]{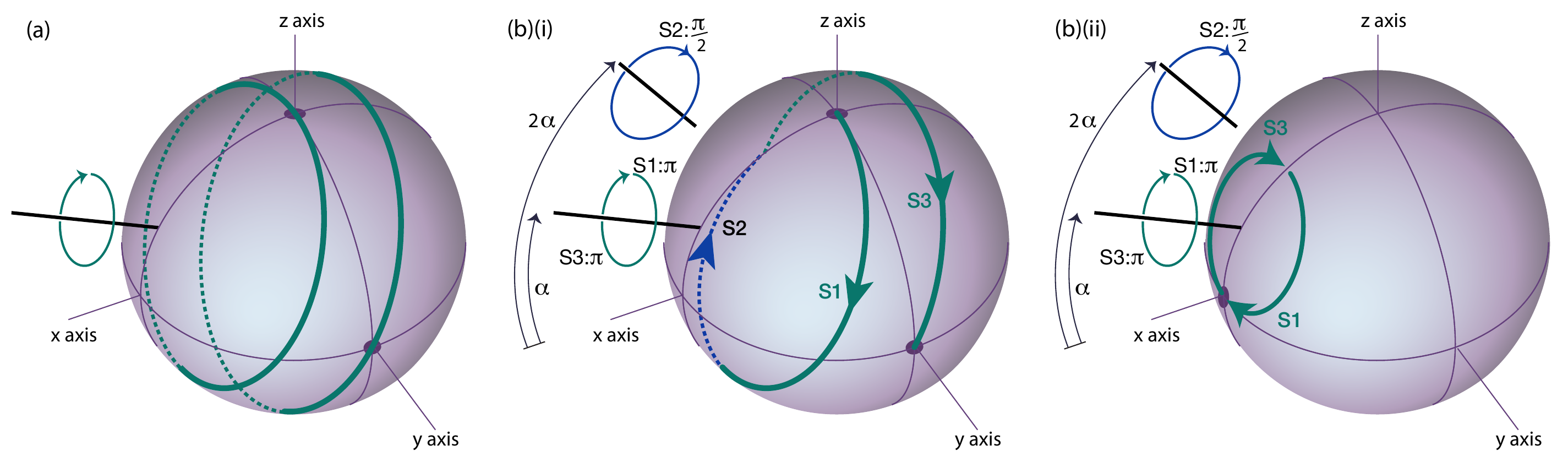}
\caption{(a) Illustration of the off-axis nature of the `natural' gate when the two dots have distinct $g$ factors: one cannot achieve a pure $x$-rotation (thus the two states marked with purple dots do not evolve into one another). (b) An illustration of the basic `ABA' pulse sequence which corrects the off-axis rotation error. The first and last of the three steps (marked green) correspond to the same rotation, by $\pi$ about an axis making angle $\alpha$ with respect to the $x$-axis. The middle step (marked blue) is a $\pi/2$ rotation about an axis which makes an angle $2\alpha$ with respect to the $x$-axis. In (b)(i) one sees that the state $\ket{01}$ at the north pole undergoes a net rotation to $(\ket{01}+i\ket{10})/\sqrt{2}$ as desired (the two filled purple dots mark these states). In (b)(ii) we see that the same sequence maps the state marked with the purple dot onto itself, only acquiring phase, since the first pulse leaves it an eigenstate of the second pulse (i.e. on that pulse's axis). Together, b(i) and b(ii) imply that we have synthesised the desired ideal rotation about the x-axis. Note that the angle $\alpha$ is exaggerated for this schematic, as compared to the value in the pulse sequence ($\alpha\approx 7.8^\circ$).}
\label{fig:bloch_sphere}
\end{figure*}

\subsection{Robust pulse sequence}
Robust pulse sequences or composite pulse sequences have been devised and implemented over several decades by the NMR community, to mitigate the effect of experimental imperfections \cite{levitt_1986}. They can also be used in the context of quantum computation to reduce the last two types of errors mentioned above (see e.g. Refs.~\cite{brown_2004, jones_2009}). 
There are two different classes of composite pulses: those that are efficient for a subset of initial states and those which are error-tolerant for any initial state (also known as fully compensating pulses). In this work we focus on a fully compensating pulse that generates a $\sqrtswap$ gate.

\begin{figure}
    \centering
    \includegraphics[width = \linewidth]{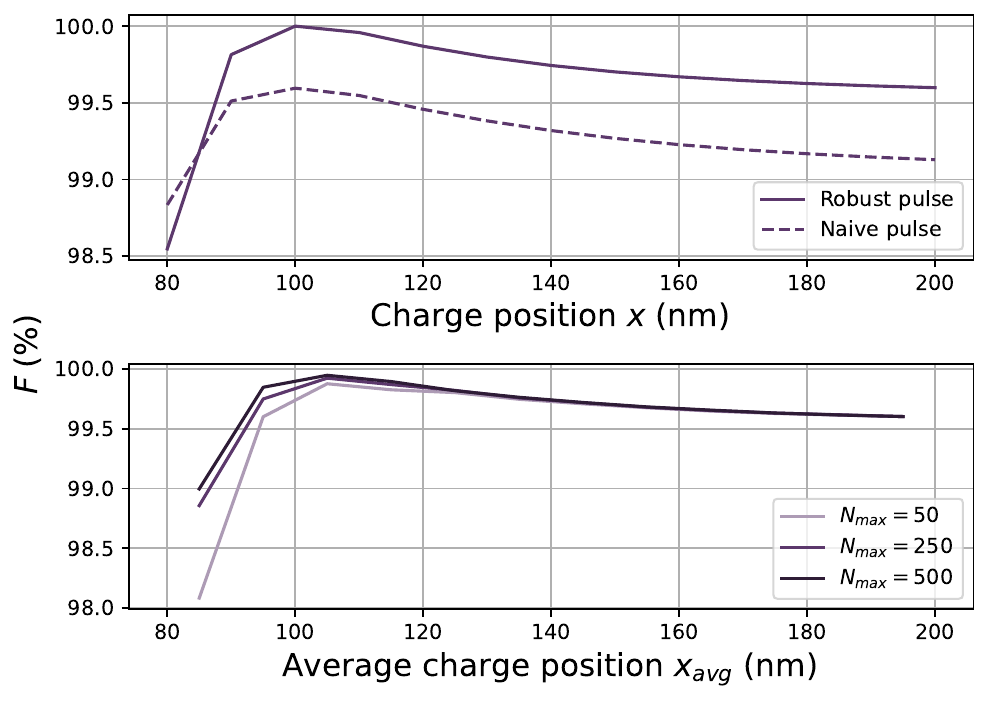}
    \caption{Evolution of the fidelity when using the robust pulse sequence for a static (top) and fluctuating charge (bottom). For the fluctuating case, the charge oscillates between two traps located at a fixed $10$ nm distance; the horizontal axis in the graph is the mean charge position. The number of fluctuations during the pulse is drawn uniformly at random between $N = 1$ and $N = N_{max}$. Additionally, the switching times are randomly chosen between $t = 0$ and $t = T_{\sqrtswap}$. Each curve is obtained by averaging $10^4$ instances of such random environment. For the naive case, we used a maximal $J$ coupling equal to the strongest one appearing in the robust pulse sequence.}
    \label{fig:pulse_sequence_fluc}
\end{figure}

This type of pulse (or pulse sequence) will generally take longer to apply than the naive one.
Hence, grid-based methods may become impractical for the search of robust pulse sequences: the large number of time steps we need to propagate the full wavefunction is compounded by the need to assess each candidate sequence against a variety of charge defect scenarios. Fortunately, we can restrict the use of full grid-based model to characterising $p_{loss}$ and providing certain key parameters which we can then use to construct a simpler model with a far smaller subspace. In our case, the subspace of interest is spanned by the states $\ket{01}$ and $\ket{10}$. In this subspace, the time-dependent Hamiltonian is given by,

\begin{align}
    H(t) = J(V_g(t))\sigma_x + \Delta E_z  \sigma_z,
\end{align}
with the dependence of $J$ on $V_g$ having been extracted from the grid-based method. Here $\Delta E_z$ is the difference in Zeeman energy between the two dots due to their difference in $g$-factor and $\sigma_z, \sigma_x$ are the standard Pauli Z and Pauli X operators respectively. We set $\Delta E_z = J_{on}/20$ (with $J_{on}$ the maximal exchange value obtained when switching the interaction on) which is a typical value for the magnetic detuning \cite{jacob_2025}. 
Since $\Delta E_z$ is assumed to be fixed and non-zero, we cannot directly perform a pure $x$-rotation for any value of our controllable parameter $J$. However different values of $J$ will result in rotations around different axes within the relevant quadrant of the $x$-$z$ plane (see Fig.\,\ref{fig:bloch_sphere}). This freedom allows us to construct interesting sequences.

The ideal composite pulse would implement a high fidelity on-axis rotation (around the x-axis), despite both the off-axis nature of the primitive evolution and the problem of charges in the environment. One should aim for a fully-compensating sequence that does so for {\em any} input state, i.e. a good match to the ideal unitary. 

We explored several potential sequences, and we report the performance of one successful option in Fig.\,\ref{fig:pulse_sequence_fluc}. This is a relatively complex sequence with 9 steps. In creating this sequence, our approach was to begin with a more simple sequence of the form $ABA$ which is well-known\,\cite{jjonesNMRnotes} to correct off-axis errors, confirming that this does so in the absence of charge noise (see Fig.\,\ref{fig:bloch_sphere}). Then, we divided this sequence into two equal parts as $(A\frac{B}{2})(\frac{B}{2}A)$, and looked for additional pulses that could be inserted into the middle in order to compensate for the charge noise effects. We adapted another solution employed in the NMR literature\,\cite{jjonesNMRnotes} whereby we rotate by $\pi$ about an axis approximately perpendicular to the $x$-$z$ plane (requiring two pulses in our case) and then perform a full $2\pi$ rotation about an axis in the $x$-$z$ plane, before reversing the earlier $\pi$ rotation. The net effect is to accumulate charge-induced rotation of opposite sign to that which occurs in the unembellished $ABA$ sequence -- thus the charge-induced errors which occur `naturally' are largely negated by those which we `deliberately' accumulate.

This specific fixed sequence is effective in a range of charge scenarios. With a numerical search, it can be further tuned (without presuming foreknowledge of the charge environment) to make it maximally effective against charge impurities at a given distance, at the cost that sequence will no longer implement a perfect gate in the case of an pristine zero-charge environment. This is the approach taken for the sequence whose performance is shown in Fig.\,(\ref{fig:pulse_sequence_fluc}).

While the sequence we have explored here does indeed boost gate fidelity, this comes at the cost of greatly increasing the total gate time. Compared to the naive pulse which allowed us to perform a $\sqrtswap$ gate in $T_{\sqrtswap} \approx 12$ ns in \cref{fig:smooth_swap}, the robust pulse sequence leads to $T_{\sqrtswap} \approx 1.9$ \textmu s. We emphasise that we did not make a systematic search for more compact sequences, and this is an interesting challenge for future work.

\subsection{Numerical results}
As we are interested in fully-compensating pulses, our measure of fidelity should now focus on comparing the implemented unitary to the target one, rather than comparing specific states. In this section we define the fidelity $F$ as \cite{pedersen_2007b},
\begin{align}
    F(U, U_{target}) \coloneqq \frac{d+\Tr(U^{\dagger}U_{target})^2}{d(d+1)},
\end{align}
with $U_{target} = U_{\sqrtswap}$ restricted to the subspace spanned by the states
$\{ \ket{01}, \ket{10}\}$ and $d = 2$.

At the top of \cref{fig:pulse_sequence_fluc}, we plot the evolution of $F$ for different positions of a static charge impurity when using the naive and robust pulses. {In this section we use a maximal $J$ coupling equal to the strongest one appearing in the robust pulse sequence for the naive pulse}. By tuning the robust pulse sequence such that the maximal fidelity is obtained when $x = 100$ nm, we obtain an increased fidelity over a large range of positions. As a consequence, the pulse cannot generate a perfect $\sqrtswap$ gate in the limit of an infinitely remote impurity (or no impurity). As the charge becomes very close to the DQD, we see that inevitably the fidelity starts to drop drastically. Broadly one would expect that the (unexpected, uncalibrated for) presence of a defect charge directly above the double dot would be catastrophic to gate fidelity regardless of mitigation efforts -- effectively that dot structure would be, at least temporarily, non-functional as a computing component. If we wished to study these scenarios more fully, a finer model of the impurity would be necessary to reach meaningful results (likely promoting the impurity to being modelled as a full quantum entity).

As the pulse sequence is two orders of magnitude longer than the naive one, we might expect to have significant number of fluctuations during the process. 
It is important to note that by switching from the grid-based method to this lower-dimension model, we lost the ability to track the loss probability. However, for the positions of the charge considered here, $p_{loss}$ is small and is not drastically modified by the charge fluctuations as reported in \cref{fig:charge_fluc_plot}.

At the bottom of \cref{fig:pulse_sequence_fluc}, we plot the evolution of $F$ for different average position of the charge $x_{avg}$ in the dot line (which is the worst-case). Having an average position $x_{avg}$ represents the situation of a charge oscillating between two traps positioned at a fixed distance of $10$ nm around $x_{avg}$. For instance, for $x_{avg} = 95$ nm, the charge oscillates between $x = 90$ and $x = 100$ nm. Given the number of fluctuations $N$ and their timings $\{t_i\}_{1 \leq i \leq n}$ are unknown, both are randomly generated such that $N$ is uniformly selected between $N = 1$ and $N = N_{max}$, while the $t_i$ are drawn randomly between $t = 0$ and $t = T_{\sqrtswap}$. Each curve is the result of averaging over $10^4$ instances of such a random environment. 

In this plot, we can distinguish two different regimes. On the one hand, when the average position of the charge is far from the DQD, the number of fluctuations has negligible impact. This aligns with the trend observed in \cref{tab:exchange_noise} -- the farther the charge is from the dot, the smaller the change in the exchange coupling. Hence, fluctuating between two positions around $x_{avg}$ that yield a similar $J$ leads to a fidelity similar to that of a fixed charge at $x_{avg}$, irrespective of the number of fluctuations. This also explains why we retrieve a similar evolution as in the static case (solid purple line on the top of \cref{fig:pulse_sequence_fluc}). 
On the other hand, when the charge oscillations occur closer to the dot, the difference in the impact on $J$ between the two traps is larger, resulting in a smaller fidelity. However, with an increasing number of fluctuations, the fidelity improves and tends to converge toward that of the static case. 

In both the static and fluctuating charge scenarios, our pulse sequence shows improvements on the fidelity over a large range of positions, thus accommodating for various charge environments. We believe this demonstrates the the modelling approach taken here can readily be used to seek and validate robust pulses, potentially allowing for the discovery of various sequences which yield high-fidelity gates. 

\section{\label{sec: conclusion} Future directions and conclusion}

In this work we present a versatile modelling approach able to simulate electrons and their spins in double quantum dots. Using realistic potentials, the tools we employ are able to identify promising devices by computing the exchange coupling, an important metric in the realisation of two-qubit gates. Moreover, the tools can fully simulate the qubits' dynamics and allow one to study various voltage control pulses in order to increase gate fidelities as well as reducing (highly damaging) risk of losing qubits.

Using a two-level fluctuator model, we explored different scenarios in which a single charge trapped near the DQD impacts the quality of the gate. For a static charge, we found that the fidelity can be quite heavily impacted whereas the loss stays constant (provided a suitable adiabatic control pulse is used). However, for a randomly oscillating charge we have both an impact on the fidelity and on the loss. While the impact on the fidelity can be mitigated using robust pulse sequences, as shown for a reduced model, the effect on the loss cannot be removed for unknown environments. 

The versatility of the grid-based methods opens many other future directions. Our study demonstrated that a quasi-2D method can near-perfectly replicate the results of a far more numerical costly 3D solution. It would be interesting to explore this for cases where the $z$-confinement is irregular, making the mapping to 2D more challenging. The complexity of the Hamiltonian could also be increased to incorporate more physical phenomena; expanding our model by adding spin-orbit interaction terms will be interesting to study. The same is true for adding the valley degree of freedom and valley-orbit coupling terms. With the valley degree of freedom incorporated, increasing the model's resolution to capture interface roughness (at the interface where the electrons are confined) would be very interesting. Finally, the introduction of interactions between two-level fluctuators or the modification of the model to treat them as distinct quantum particles are other potential areas for investigation.

To conclude, our approach and the tools we have developed set the stage for a software pipeline that takes as input a chip layout defined in software used to design traditional chips, and returns two-qubit gate performances for realistic noise environments. It is worth noting that we haven't investigated single qubit gates nor other types of two-qubit gates in this manuscript, but it would be a natural extension of the methods here to do so. 

\section*{Acknowledgments}
The authors are very grateful to James Williams for help throughout this research and in particular for generating the electrical potentials, derived from electrode structures, which we use in this work. We also thank Hans Hon Sang Chan, Andrew Fisher, Guido Burkard, Tyson Jones and Richard Meister for helpful discussions.
The numerical modelling involved in this study made
use of the Quantum Exact Simulation Toolkit (QuEST)~\cite{tyson_2019}, and the recent development
pyQuEST~\cite{pyquest} which permits the user to use Python as the interface front end.
We are grateful to those who have contributed
to all of these valuable tools. 
The authors would like to acknowledge the use of the University of Oxford Advanced Research Computing (ARC)
facility~\cite{oxford_arc} in carrying out this work
and specifically the facilities made available from the EPSRC QCS Hub grant (agreement No. EP/T001062/1). The authors also acknowledge support from EPSRC’s Robust and Reliable Quantum Computing (RoaRQ) project (EP/W032635/1), and the SEEQA project (EP/Y004655/1).

\section*{Author contributions}
H.J. developed the modelling tool, performed the numerical simulations and wrote the manuscript. S.C.B. contributed to developing the method and to writing the manuscript, and wrote code to independently validate the results.

\appendix

\section{\label{app:device_spec} Device specifications}
For the device we consider (see \cref{fig:main_figure}), we used gates and oxide dimensions consistent with designs and expected fabricated dimensions from the IMEC SiMOS process. The silicon substrate is separated from the three gate layers by $8$ nm thick silicon dioxide. Each gate layer is modelled as $30$ nm thick Schottky contacts with a work function of $4.1$ eV consistent with n++ doped polysilicon with $5$ nm silicon dioxide separating each layer.
The first layer (the closest to the silicon surface) is used to define the channel through confinement gates with a separation of $64$ nm. The second layer is used for plunger gates to form our potential wells, they are separated by $30$ nm and have a length of $54$ nm (due to the oxide layer between gate layers) and width of $40$ nm making an area of $2160$ nm$^2$. The third layer incorporate source and drain gates which yields lateral confinement to our channel, and a barrier gate which sits between both plunger gates, with length $44$ nm and width of $20$ nm.
The plunger gates were set to $0.3$ V and the barrier was swept from $-0.1$ to $0.4$ V, all other gates were set to $0$ V. The potential landscape 2D slice was taken $2$ nm below the Si/SiO2 interface.

\section{\label{app:abrupt_transition} Abrupt transition}
\begin{figure}[hbt!]
    \centering
    \includegraphics[width=\linewidth]{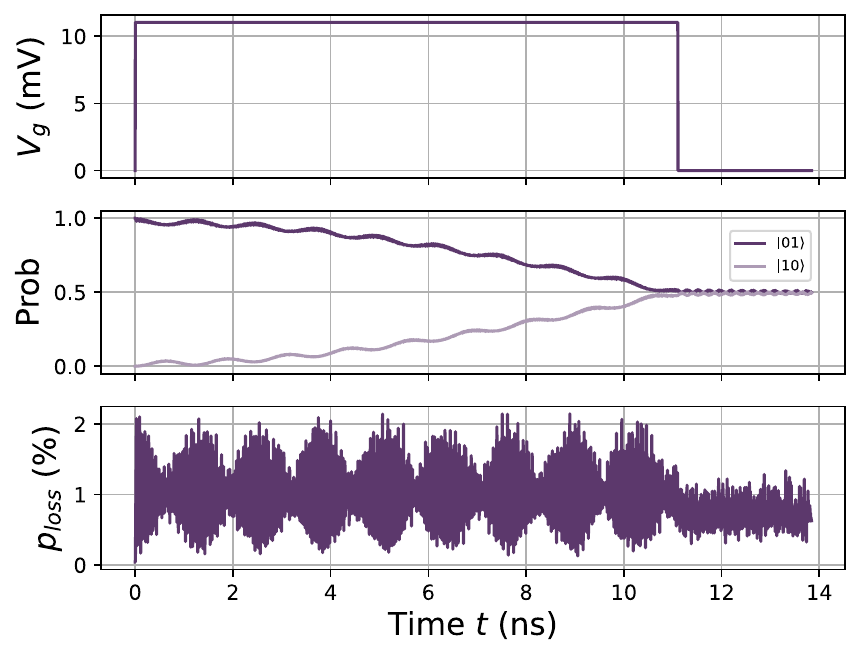}
    \caption{Abrupt $\sqrtswap$ pulse. We abruptly switch the exchange interaction on and off and observe beatings due to the excitation of higher energy charge states. As expected the loss probability is highly increased with this method and can be as large as $1\%$.}
    \label{fig:swap_abrupt}
\end{figure}

The grid-based method allows us to study any type of evolution of the potential in time. In the main text we focus on the more realistic case in which we smoothly lower and raise the barrier to perform a $\sqrtswap$ gate. The intuition behind this choice being that by adiabatically modifying the potential we will implement a more performant two-qubit gate. To validate this choice, we compare it to the extreme case in which we abruptly switch the coupling on. In \cref{fig:swap_abrupt}, we see that by doing so we largely increase $p_{loss}$ (up to $1\%$). More precisely, by taking the average over the time remaining after the gate we find that $p_{loss} = 0.73 \%$ and $F = 97.06 \%$. This result emphasises the need for careful considerations of the control pulse.

\section{\label{app:y_dir_charge} Charge trapped outside of the dot line}
As said in the main body, we are not limited to the study of charges in the dot line: we can position them anywhere we want. In this subsection, we look at the performance of the $\sqrtswap$ gate for three different position of a single charge. To make a fair comparison, we will position it at a fixed distance of $100$ nm from the DQD but at different angles, namely $0$ (in the dot line), $45$ and $90$ degrees. 

\begin{figure}[hbt!]
    \centering
    \includegraphics[width=\linewidth]{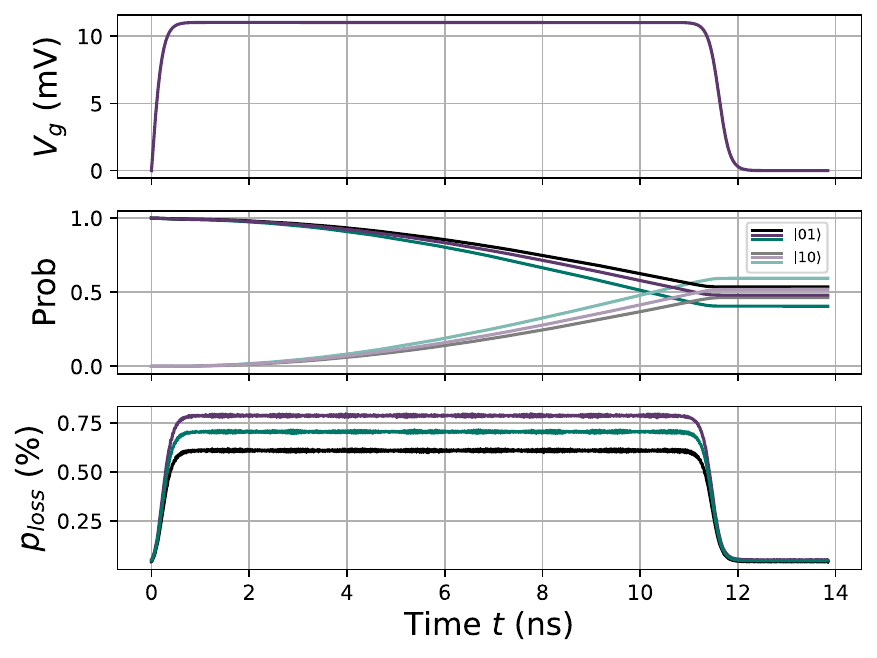}
    \caption{Smooth $\sqrtswap$ gate for charges respectively positioned at $(100, 0, 2)$ nm (green), $(0, 100, 2)$ nm (black) and $(50\sqrt{2}, 50\sqrt{2}, 2)$ nm (purple).}
    \label{fig:swap_noisy_y}
\end{figure}

In \cref{fig:swap_noisy_y}, we plot the evolution of the qubits' probability as well as $p_{loss}$ for a given voltage profile (optimised for the noiseless case) and report the final fidelities and loss probabilities in \cref{tab:swap_noisy_y}. The results correspond to our intuition as the worst fidelity is attained for a charge in the dot line. It is worth noting that when the charge is place perpendicularly to the dot line (black), the probability to find the electrons in the $\ket{01}$ and $\ket{10}$ don't cross for this voltage profile contrary to the previous charge environments studied in the paper. This is due to the fact that by placing the charge as such, we are pushing the trapped electrons apart from one another rather than bringing them closer and hence reduce the exchange coupling. 

\begin{table}[]
    \centering
    \begin{tabular}{c|c c c}
         \hline
         \hline
         Charge position (°) & 0 & 45 & 90\\
         \hline
         $p_{loss} \; (\%)$ & 0.05 & 0.05 & 0.04 \\
         $F \; (\%)$ & 99.10 & 99.96 & 99.87 \\
         \hline
         \hline
    \end{tabular}
    \caption{Loss probability and fidelity for different charge position. Each charge is placed at a fixed distance of $100$ nm from the center of the DQD. Their position can thus be characterised by their angle with respect to the dot line ($0$° corresponds to having a charge in the dot line). }
    \label{tab:swap_noisy_y}
\end{table}

\section{\label{app:charge_fluc} Fluctuating charges}

\begin{figure*}
    \centering
    \includegraphics[width=\linewidth]{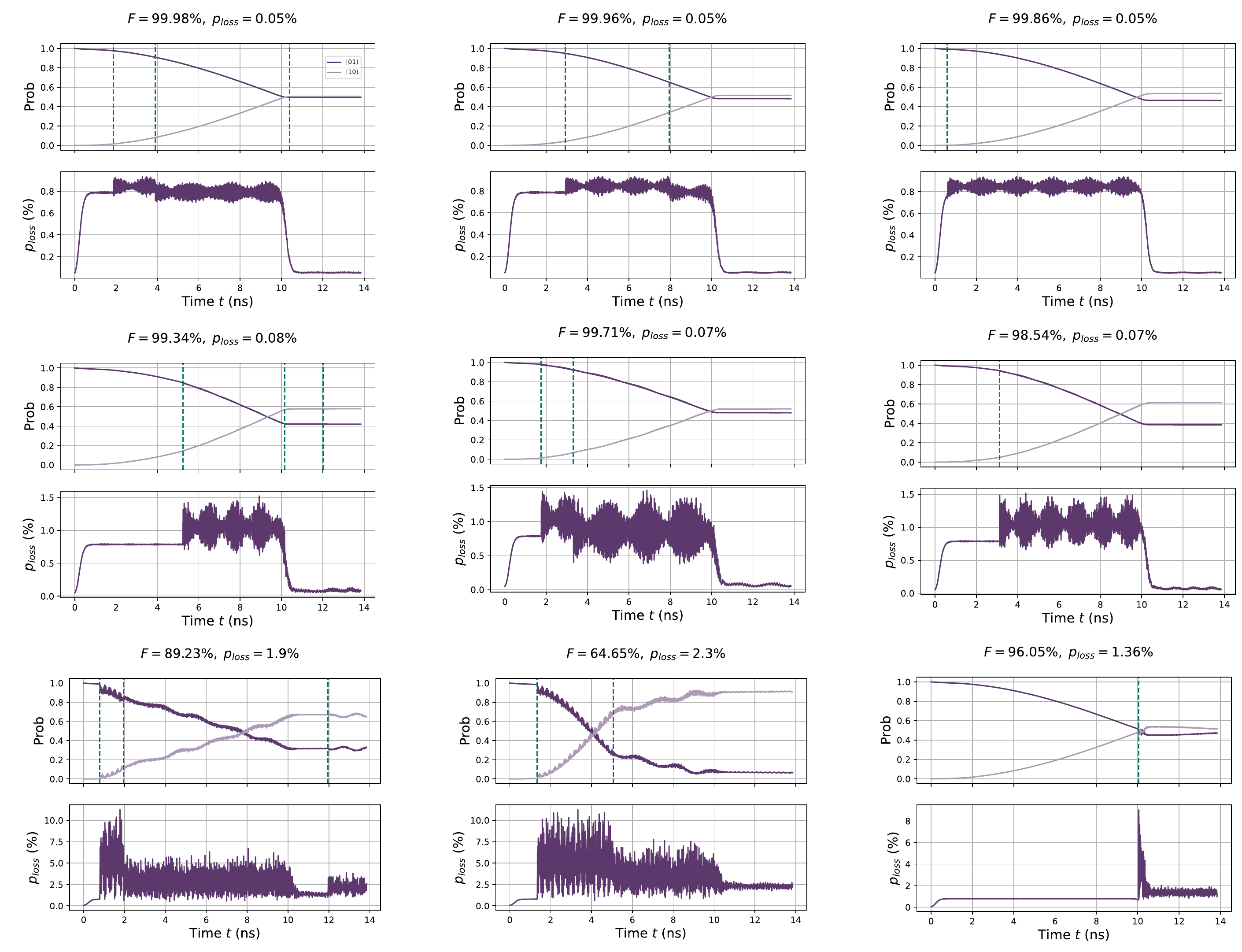}
    \caption{Evolution of the probability to find the qubits in the states $\ket{01}$ and $\ket{10}$ and the loss probability for different fluctuations scenario. Initially the charge is located at $(100, 0, 2)$ nm and switches to another position at each green dashed lines. The voltage profile has been optimised for the $(100, 0, 2)$ nm case to yield a perfect $\sqrtswap$ gate. Oscillation between $(100, 0, 2)$ nm and $(90, 0, 2)$ nm at the top, $(100, 0, 2)$ nm and $(75, 0, 2)$ nm in the middle (bottom) $(100, 0, 2)$ nm and $(50, 0, 2)$ nm at the bottom. The fidelity and loss probability are obtained by averaging over the time following the end of the gate.}
    \label{fig:charge_fluc_plot}
\end{figure*}

In this section we explore various scenarios involving multiple charge fluctuations between different positions, quantifying their impact on fidelity and loss probability, as illustrated in \cref{fig:charge_fluc_plot}. Employing a similar voltage control profile to that used \cref{fig:fluc_example}, we start with a charge placed at $100$ nm of the DQD in the dot line, leading to an initial $p_{loss} = 0.05\%$. We then perform a $\sqrtswap$ pulse and let the charge oscillates between this initial position and an additional charge trap located at $90$ nm, $75$ nm, and $50$ nm at the top, middle, and bottom of \cref{fig:charge_fluc_plot}, respectively.

We find that the impact on fidelity is similar to that of the static case, with reductions attributed to timing errors, and the magnitude of the impact amplifying as the charge approaches the DQD.
Notably, a significant departure from the static scenario lies in the loss of adiabaticity during the gate operation due to the dynamic potential changes. This is evident in the evolution of $p_{loss}$, exhibiting pronounced shifts whenever a fluctuation occurs. Consequently, there is a significant increase in $p_{loss}$ at the end of the gate, constituting the most detrimental type of error as it cannot be corrected. Similar to fidelity, the closer the charge is to the DQD the more damaged the gate's performance. For oscillations between $100$ and $50$ nm, loss probabilities reach the order of the percent, rendering the gate impractical for use.

\section{\label{app:convergence} Convergence versus basis size}

\begin{figure}
    \centering
    \includegraphics[width=\linewidth]{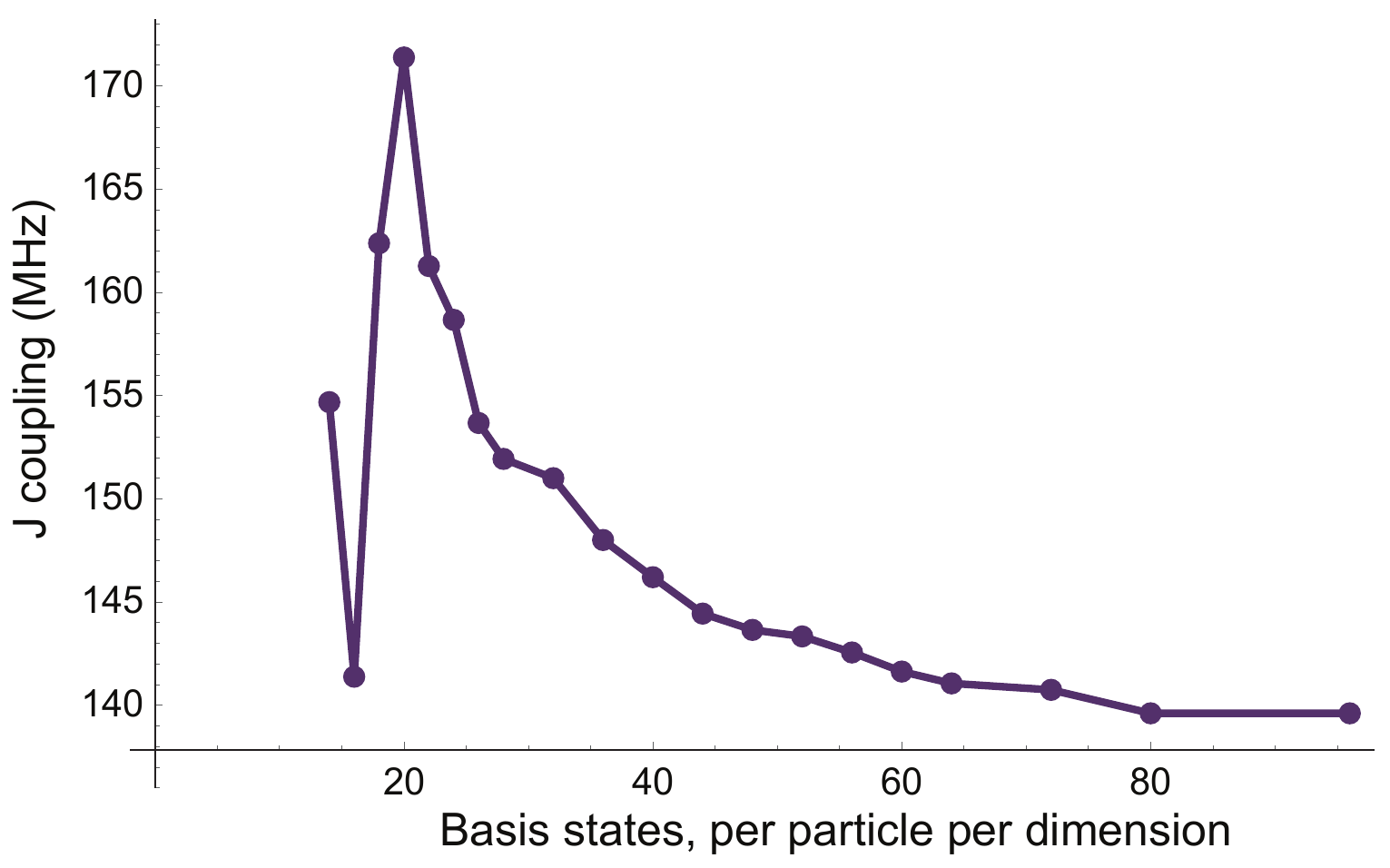}
    \caption{The J coupling strength obtained from a series of simulations with various basis set sizes, ranging from 14 to 96. Here we chose $V_g=12$\,mV. One observes that the series converges to 140\,MHz. The model used here is the variant with strict 2D and the standard Coulomb matrix, i.e. the purple lines in Fig.\,\ref{fig:exchange_couplingA}. }
    \label{fig:JversusBasis}
\end{figure}

In the main paper, the inset in Fig.\,\ref{fig:exchange_couplingA}(a) shows how the $J(V_g)$ curve can vary with the number of basis states in the model. While the general exponential dependence on $V_g$ is observed when using 16 basis states (per particle, per dimension, total 4,096), the specific values can deviate from those obtained with higher resolutions. In contrast, there is very little difference between the basis sizes of 32 (total $1$\,million) and 64 (total $16$\,millions). Given the computational resource costs it is obviously interesting not only to confirm convergence but to identify the basis size that is adequate. Therefore we coded an implementation of our model with a variable basis size, and tested a series of choices -- the results are shown in Fig.\,\ref{fig:JversusBasis}. One sees that above a basis size of around 22, the predicted $J$ values converge steadily to about $140$\,MHz. Below this size, the prediction is unstable while remaining within about $20\%$ of the converged value.

\section{\label{app: coulomb}Coulomb interaction in real space}

As explained in the main paper, the SO modelling method that we employ involves two representations for the system's state: The plane wave (or `k-space') representation where each stored amplitude corresponds to a component of the form $\exp(i 2\pi\bm k_i\cdot\bm r)$, and a second dual basis that results from a Fourier transform.

The confining potential $H_\text{conf}=V(\bm r_1)+V(\bm r_2)$, including any effects of unwanted charges in the local environment of the qubit, depends only on the real-space coordinates and so we apply the corresponding unitary while we are in the real-space basis. Because the basis functions are not perfectly local (i.e. they are not Dirac delta functions for finite $m$) it follows that strictly speaking, the matrix representing $\exp(i\,H_\text{conf}\,\delta t)$ in the Dirichlet basis is not perfectly diagonal. However the potential creating the dot confinement, due to electrode voltages, materials properties, and unwanted charges outside the primary dot region, is smoothly varying and does not change dramatically over the length scale of the grid point separations. In the present study we neglect these off-diagonal components and take the on-diagonal components as simply 
\[
\bra{n_{1,x} n_{1,y}n_{2,x} n_{2,y}}H_\text{conf}\ket{n_{1,x} n_{1,y}n_{2,x} n_{2,y}}=V(\bm{r}_1)+V(\bm{r}_2)
\]
where $V(\bm{r})$ is as defined in \cref{eq: hamiltonian} and 
\[
\bm{r}_1=S(p_{n_{x,1}}\bm i+p_{n_{y,1}}\bm j),\ \ \ \ \ 
\bm{r}_2=S(p_{n_{x,2}}\bm i+p_{n_{y,2}}\bm j).
\]
Presently we discuss the inclusion of a finite $z$-depth; this is also neglected in the treatment of $H_\text{conf}$ since the potentials vary little over the modest (order 5\,nm) depth.

The Coulomb interaction between the two electrons (the two qubits) in our model should be considered in more detail: it is singular and this is one of the challenges of modelling multi-electron systems. Indeed the implications of the singularity are so strong that in one dimension the model has pathologies such as a singular ground state~\cite{Loudon2016}. Fortunately in two-and-higher dimensions these states are well-behaved. In the remainder of this appendix section we will
\begin{itemize}
\item derive the expression for the matrix elements of $H_\text{coul}$ in the real-space Dirichlet basis, both for strictly 2D and for `pizza box' type 3D potentials;
\item determine the numerical values of the on-diagonal elements for relevant cases;
\item discuss the strength of the off-diagonal terms and how future work can include them in the simulation as a correction.
\end{itemize}

To keep the expressions in this section as clean as possible we will specify that the spatial `box' within which our electrons are represented runs from $-\frac{1}{2}$ to $+\frac{1}{2}$ on both the $x$ and $y$ axes, and therefore has unit area. Obviously, the expressions we obtain can be scaled to arbitrary sizes; moreover the generalisation to e.g. elongated boxes is straightforward.

\subsection{2D quantum dots}

Because our basis is periodic (i.e. the wavefunctions simply repeat in $x$ and $y$ with a period of unity), we should adopt a periodic function in place of the canonical non-periodic Coulomb interaction. 
 Writing the vector between the two particles as $\Delta \equiv \bm r_1 - \bm r_2 $, when $|\bm i\cdot\Delta\bm r|=\frac{1}{2}$, then particle 1 is equidistant from its real partner particle 2 and an image of that particle -- our choice should acknowledge that symmetry (and similarly in the $y$ direction). A convenient choice is simply to use the canonical Coulomb interaction in the region near the origin and then repeat it outside that region:
 \begin{eqnarray}
 &&C(\Delta) = \frac{1}{|\Delta|}
 \ \ \ \text{when} \ \ \ 
 |\bm i\cdot\Delta|\leq\frac{1}{2}, \ \
 |\bm j\cdot\Delta|\leq\frac{1}{2}\nonumber\\
 &&C(\Delta +a\,\bm i +b\,\bm j) = C(\Delta )\ \ \ \ \forall\, a,b\in \mathbb{Z} 
\label{eqn:periodicC}
\end{eqnarray}
This choice is appropriate for our scenario where the image charges are an artifact of the simulation and do not reflect any element of physical reality (i.e. the electrons in the dot structure do not actually experience image charge effects). We must ensure the simulation box is large enough that the double-dot confinement region is only in the central part of the simulation box, so that there is negligible amplitude associated with states for which $|\bm i\cdot\Delta|>\frac{1}{2}$ or $|\bm j\cdot\Delta|>\frac{1}{2}$ (as such states experience the `wrong' Coulomb interaction). In the main paper we report that we verify this both by tracking the total probability outside of the central region, and by doubling the simulation box size to ensure that properties of interest to not change (convergence with respect to box size). 

We can then write the general matrix element as 
\begin{eqnarray}
&&\bra{n^\prime_{1,x} n^\prime_{1,y} n^\prime_{2,x} n^\prime_{2,y}}H_\text{coul}\ket{n_{1,x} n_{1,y}n_{2,x} n_{2,y}}=\quad\quad \nonumber\\
&&\ \ 
K\!\!\int\displaylimits_\text{box}\!\! d\bm r_1\!\!\int\displaylimits_\text{box}\!\! d\bm r_2\ 
C(|\bm r_1-\bm r_2|)\ 
\phi_{n^\prime_{1,x}\,n^\prime_{1,y}}^*\!(\bm r_1)\ \phi_{n^\prime_{2,x}\,n^\prime_{2,y}}^*\!(\bm r_2)\nonumber\\
&&
\qquad\qquad\qquad\qquad \times\phi_{n_{1,x}\,n_{1,y}}\!(\bm r_1)\ 
\phi_{n_{2,x}\,n_{2,y}}\!(\bm r_2)\ \ \ 
\label{eqn:fullMatrixEle}
\end{eqnarray}
where $K$ is the appropriate constant incorporating $e^2/4\pi\epsilon_0\epsilon_r$ and a factor accounting for the physical width of our simulation box. From \cref{eqn:RSpixel} we know
\begin{eqnarray}
&&\phi_{a\,b}(\bm r)=\frac{1}{M}
\exp(i \pi (x\!-\!\frac{a}{M}\ +\ y\!-\!\frac{b}{M}))\\
&&\qquad\qquad\qquad\quad \times D_m(2\pi(x-\frac{a}{M}))D_m(2\pi(y-\frac{b}{M}))\nonumber
\label{EqnPhiForm}
\end{eqnarray}
recalling that $M=2^q$ is the total number of basis states associated with each dimension for each particle, and $m=M/2$. 

Before proceeding we note that because of the transnational symmetry in the expression \cref{eqn:fullMatrixEle} the value of the matrix element will be unchanged if we subtract any integer from all the $x$ quantum numbers, and/or subtract any integer from all the $y$ quantum numbers (wrapping around the cycle $-m$ to $m-1$ as needed). Therefore we need only solve for the cases where $n_{1,x}=n_{1,y}=0$ since the elements for non-zero  $n_{1,x}$, $n_{1,y}$ can always be found  by suitable shifts.

We now make the change of variable,
\[
\bm S=\bm r_1+\bm r_2\quad\quad
\bm \Delta=\bm r_1-\bm r_2.
\]
Noting from the Jacobian that $d\bm r_1\,d\bm r_2=\frac{1}{4}d\bm S\, d\bm \Delta$, the integral in \cref{eqn:fullMatrixEle} would directly translate to 
\[
\int\displaylimits_\text{box}\!\! d\bm r_1\!\!\int\displaylimits_\text{box}\!\! d\bm r_2\ 
\rightarrow
\frac{1}{4}
\int\displaylimits_\textsc{BOX}\!\! d\bm \Delta\!\!\int\displaylimits_{\text{rect}(\bm \Delta)}\!\!\!\! d\bm S 
\]
where the $\bm \Delta$ integral is over a region denoted `\small{BOX}' in capitals as it centred at the origin but twice the linear size (four times the area) of the `box' regions, i.e. the region $|\bm i\cdot\bm \Delta|\leq 1$ and $|\bm j\cdot\bm \Delta|\leq 1$. Meanwhile the $\bm S$ integral is over a region denoted $\text{rect}(\bm \Delta)$ which depends on $\bm \Delta$:
\[
|\bm i\cdot\bm S|\leq
1-|\bm i\cdot\bm \Delta|
\qquad
|\bm j\cdot\bm S|\leq
1-|\bm j\cdot\bm \Delta|.
\]
This constraint ensures that every $\bm r_1$,  $\bm r_2$ point in the original integral is included, but not points outside that space.

\begin{figure}
    \centering
    \includegraphics[width=0.49\textwidth]{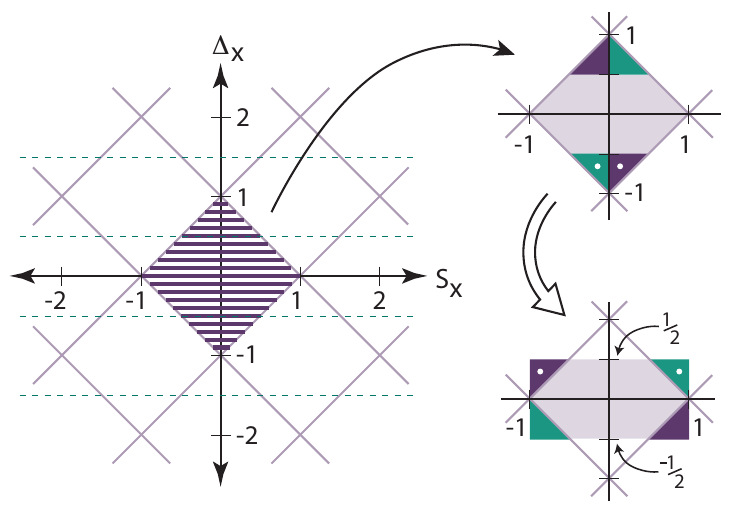}
    \caption{The figure illustrates equivalent options for integration; the $x$-coordinates are shown and an identical argument applies, independently, to the $y$-coordinates. Left side: The wavefunctions in \cref{eqn:fullMatrixEle} are periodic, and one cell  (i.e. the `box') corresponds to sweeping $r_{1,x}$ and $r_{2,x}$ over the range $-\frac{1}{2}$ to $\frac{1}{2}$. Correspondingly the $\Delta_x$ and $S_x$ variables sweep the region shaded in dark blue, and the periodicity of the wavefunctions in these variables is indicated by the cyan diagonals.  The red dashed lines indicate the periodicity of  the other factor in the integrand, $C(|\Delta|)$. Right side: Because of the indicated periodicities, we know that each triangle of a given colour in the upper figure is equivalent to the same triangle in the lower figure. Consequently we can simplify the limits of the integral as in the main text. }
    \label{fig:periodicity}
\end{figure}

One could continue with the following steps using the $\Delta$-dependent integration region\footnote{as a check, this was done to ensure the numerics come to the same value}. However it is convenient to observe that,
because of the periodicity of the integrand, 
\[
\int\displaylimits_\textsc{BOX}\!\! d\bm \Delta\!\!\int\displaylimits_{\text{rect}(\bm \Delta)}\!\!\!\! d\bm S 
\rightarrow
\int\displaylimits_\text{box}\!\!\! d\bm \Delta\!\!\int\displaylimits_{\text{BOX}}\!\!\! d\bm S 
\]
where the region of the $\bm \Delta$ integral becomes the unit-area box centred at the origin, i.e. $|\bm i\cdot\bm S|\leq \frac{1}{2}$ and $|\bm j\cdot\bm S|\leq \frac{1}{2}$, and the region of the $\bm S$ integral is becomes larger \small{BOX} referred to above. See Fig.~\ref{fig:periodicity} for an illustration of the equivalence. 

With these observations we are left wishing to evaluate,
\begin{eqnarray}
&&\bra{n^\prime_{1,x} n^\prime_{1,y} n^\prime_{2,x} n^\prime_{2,y}}H_\text{coul}\ket{0\ 0\ n_{2,x} n_{2,y}}=\quad\qquad\qquad \qquad\qquad\nonumber\\
&&\qquad\qquad\qquad\qquad\qquad\quad
\frac{1}{4}
\int\displaylimits_\text{box}\!\!\! d\bm \Delta\ 
C(|\bm \Delta|)\,f(\bm \Delta)
\label{eqn:DeltaAlone}
\end{eqnarray}
where
\begin{eqnarray}
f(\bm \Delta)&=& \int\displaylimits_{\text{BOX}}\!\!\! d\bm S\,\phi_{n^\prime_{1,x}\,n^\prime_{1,y}}^*\!(\frac{\bm S + \bm \Delta}{2})\ \phi_{n^\prime_{2,x}\,n^\prime_{2,y}}^*\!(\frac{\bm S - \bm \Delta}{2})\qquad\qquad\nonumber\\
&\,&
\qquad\qquad \times\phi_{0\,0\,}\!(\frac{\bm S + \bm \Delta}{2})\ 
\phi_{n_{2,x}\,n_{2,y}}\!(\frac{\bm S - \bm \Delta}{2})\nonumber\\
\label{eqn:introducingf}
\end{eqnarray}

The $f(\bm \Delta)$ can be simplified, first by recalling from \cref{EqnPhiForm} that our basis states factor into $x$- and $y$-
dependent functions, so that 

\begin{eqnarray}
f(\bm \Delta) &=& f(\Delta_x,\Delta_y)=\frac{e^{i\Theta}}{M^4}\ g_x(\Delta_x)\ g_y(\Delta_y)
\\
\text{where}\ \Theta &=& n^\prime_{1,x} +
 n^\prime_{1,y} + n^\prime_{2,x} + n^\prime_{2,y} - n_{2,x}- n_{2,y}\ \ \ \nonumber
 \label{eqn:fAndTHETA}
\end{eqnarray}
and 
\begin{eqnarray}
&&g_x( \Delta)=\nonumber\\
&&\int\displaylimits_{-1}^{+1}\!\! d S\,
D_m(2\pi(\frac{ S + \Delta}{2}-\frac{n^\prime_{1,x}}{M}))\ 
D_m(2\pi(\frac{ S - \Delta}{2}-\frac{n^\prime_{2,x}}{M}))\nonumber\\
&&
\qquad \times D_m(2\pi(\frac{ S + \Delta}{2}))\ 
D_m(2\pi(\frac{ S - \Delta}{2}-\frac{n_{2,x}}{M}))
\label{eqn:introducingf}
\end{eqnarray}
and similarly for $g_y(\Delta)$.

Finally we note that we can subsume the quantum number $n_{2,x}$ into a new integer $n^{\prime\prime}_{2,x}=n^\prime_{2,x}-n_{2,x}$, reducing the number of parameters, as follows. We examine $g_x(\Delta-\frac{n_{2,x}}{M})$ and shift the integration variable $S\rightarrow S-\frac{n_{2,x}}{M}$, (which does not require an adjustment to the integration range, since that range corresponds to one period of the periodic integrand). Then we find
\[
g_x( \Delta-\frac{n_{2,x}}{M})=h_{n^\prime_{1,x}\,n^{\prime\prime}_{2,x}}(\Delta)
\]
where
\begin{eqnarray}
&&h_{a\,b}(\Delta)=\nonumber\\
&&\int\displaylimits_{-1}^{+1}\!\! d S\,
D_m(2\pi(\frac{ S + \Delta}{2}-\frac{a}{M}))\ 
D_m(2\pi(\frac{ S - \Delta}{2}-\frac{b}{M}))\nonumber\\
&&
\qquad \times D_m(2\pi(\frac{ S + \Delta}{2}))\ 
D_m(2\pi(\frac{ S - \Delta}{2}))
\label{eqn:introducingh}
\end{eqnarray}
Then we can write
\begin{eqnarray}
&&\!\!\!\!f(\Delta_x\bm i+\Delta_y\bm j)=\\
&&\frac{e^{i\Theta}}{M^4}
h_{n^\prime_{1,x}\,n^{\prime\prime}_{2,x}}(\Delta_x+\frac{n_{2,x}}{M})\ h_{n^\prime_{1,y}\,n^{\prime\prime}_{2,y}}(\Delta_y+\frac{n_{2,y}}{M}))\nonumber
\end{eqnarray}
This is a useful form since it depends only on $h_{a\,b}(\Delta)$ which can be pre-computed as a high-granularity lookup table for each required $a$, $b$. For the on-diagonal matrix elements, $a=b=0$, so that we need only pre-compute $h()$ once for a given choice of basis size $M$ in order to obtain the diagonal approximation.

Given a pre-computed $h()$ function, we still need to perform the outer $d\bm \Delta$ integral from 
Equation (\ref{eqn:DeltaAlone}). This is conveniently performed in polar variables,
\begin{eqnarray}
&&\frac{1}{4}
\int\displaylimits_\textsc{box}\!\!\! d\bm \Delta\ 
C(|\bm \Delta|)\,f(\bm \Delta)=\nonumber\\
&&\frac{1}{4}
\int\displaylimits_0^{2\pi}\!\! d\theta \int\displaylimits_0^{\Delta_\text{max}(\theta)}
\!\!\!\!\!\! \Delta\,d\Delta\ C(\Delta)\,f(\bm i \Delta\cos\theta+\bm j \Delta\sin\theta)=\nonumber\\
&&\frac{e^{i\Theta}}{2M^4}
\int\displaylimits_0^{2\pi}\!\! d\theta \int\displaylimits_0^{\Delta_\text{max}(\theta)}
\!\!\!\!\!\! d\Delta\ h_{n^\prime_{1,x}\,n^{\prime\prime}_{2,x}}(\Delta\cos\theta+\frac{n_{2,x}}{M})\nonumber\\ 
&&\qquad\qquad\qquad\qquad h_{n^\prime_{1,y}\,n^{\prime\prime}_{2,y}}(\Delta\sin\theta+\frac{n_{2,y}}{M}).
\label{eqn:polar}
\end{eqnarray}
In the second line we were able to remove the singularity by cancelling the $\Delta$ introduced from the area element, with $C(\Delta)=1/\Delta$. The limit $\Delta_\text{max}$ of the radial integral ensures that we sweep over our square, unit-area `\small{box}' area:
\begin{eqnarray}
\Delta_\text{max}(\theta)
&=&\frac{1}{2}|\text{cosec}(\theta)|\ \text{if}\ \  \frac{\pi}{4}\!<\!\theta\!<\!\frac{3\pi}{4}\ \  \text{or}\ \ \frac{5\pi}{4}\!<\!\theta\!<\!\frac{7\pi}{4}\nonumber\\
&=&\frac{1}{2}|\text{sec}(\theta)|\ \  \text{otherwise.}
\end{eqnarray}

To summarize the analysis for the 2D problem, we find that the matrix elements of the Coulomb interaction $1/|\bm r_1-\bm r_2|$ can be computed using
\begin{eqnarray}
&&\bra{n^\prime_{1,x} n^\prime_{1,y} n^\prime_{2,x} n^\prime_{2,y}}H_\text{coul}\ket{0\,0\,n_{2,x} n_{2,y}}=\quad \nonumber\\
&&K\frac{e^{i\Theta}}{4M^4}
\int\displaylimits_0^{2\pi}\!\! d\theta \int\displaylimits_0^{\Delta_\text{max}(\theta)}
\!\!\!\!\!\! d\Delta\ h_{n^\prime_{1,x}\,n^{\prime\prime}_{2,x}}(\Delta\cos\theta+\frac{n_{2,x}}{M})\nonumber\\ 
&&\qquad\qquad\qquad\qquad h_{n^\prime_{1,y}\,n^{\prime\prime}_{2,y}}(\Delta\sin\theta+\frac{n_{2,y}}{M})
\label{eqn:finalMatrixEle}
\end{eqnarray}
where the function $h()$ is defined in \cref{eqn:introducingh} and the relative quantum numbers $n^{\prime\prime}_{2,x}=n^\prime_{2,x}-n_{2,x}$ and $n^{\prime\prime}_{2,y}=n^\prime_{2,y}-n_{2,y}$. 
The phase $\Theta$ is as defined in \cref{eqn:fAndTHETA}.
We noted that elements for which the $n_{1,x}$, $n_{1,y}$ quantum numbers are not zero, can be found simply by shifting.

With these expressions we can calculate any matrix element; of particular interest is the element $\bra{0,0,0,0}H_\text{coul}\ket{0,0,0,0}$ which is the interaction for two electrons sharing the same basis state, i.e. the `shared grid point' case. This will be the matrix element with the greatest magnitude; the state's amplitude for such an element will be zero for antisymmetric charge states, but can be non-zero for symmetric charge states. The calculated values for three basis sizes are shown in the $z_\Delta=0$ column of Table\,\ref{tab:coulombTerms}.

\subsection{3D quantum dots}

We consider a straightforward generalisation to describe `pancake' three-dimensional quantum dots where the vertical confinement is far more severe than the $x$-$y$ potential. We then make the simple assumption that the electron's wavefunctions have a separable $z$ component, and that they are `frozen' in the lowest possible $z$-eigenstate. In effect, we have only a single basis state for the $z$-direction of the model. This is an acceptable approximation unless the depth of the real dot structure, more specifically the mean $z$ separation, is  significantly larger than the grid point spacing in the $x$-$y$ plane; in that case, the 3D pixels would become `cigar shaped' with the long axis vertical, yet the model could not properly account for vertical electron-electron repulsion {\em within} each such pixel.

Note that the grid-based method presented in this paper can naturally extend to fully model the third dimension with multiple basis states. In that case, the expressions in the previous section would simply be adapted so that, for example, the polar integral \cref{eqn:polar} is performed in cylindrical-polar coordinates. In terms of the resource costs of the numerical modelling of the two-electron system, introducing $n$ basis states for the $z$-direction would increase the RAM requirements of the simulation by $n^2$. In the present paper we do not explore this, but it is an interesting theme for future work.

Here we assume 
\[
\Psi_{3D}(x_1,y_1,z_1; x_2, y_2, z_2)=Z(z_1)Z(z_2)\Psi_{2D}(x_1,y_1; x_2, y_2)
\]
Here $\Psi_{2D}(...)$ is an instance of the 2D grid-based two-electron wavefunction that has been considered in the prior section. 

\begin{figure*}
    \centering
    \includegraphics[width=\textwidth]{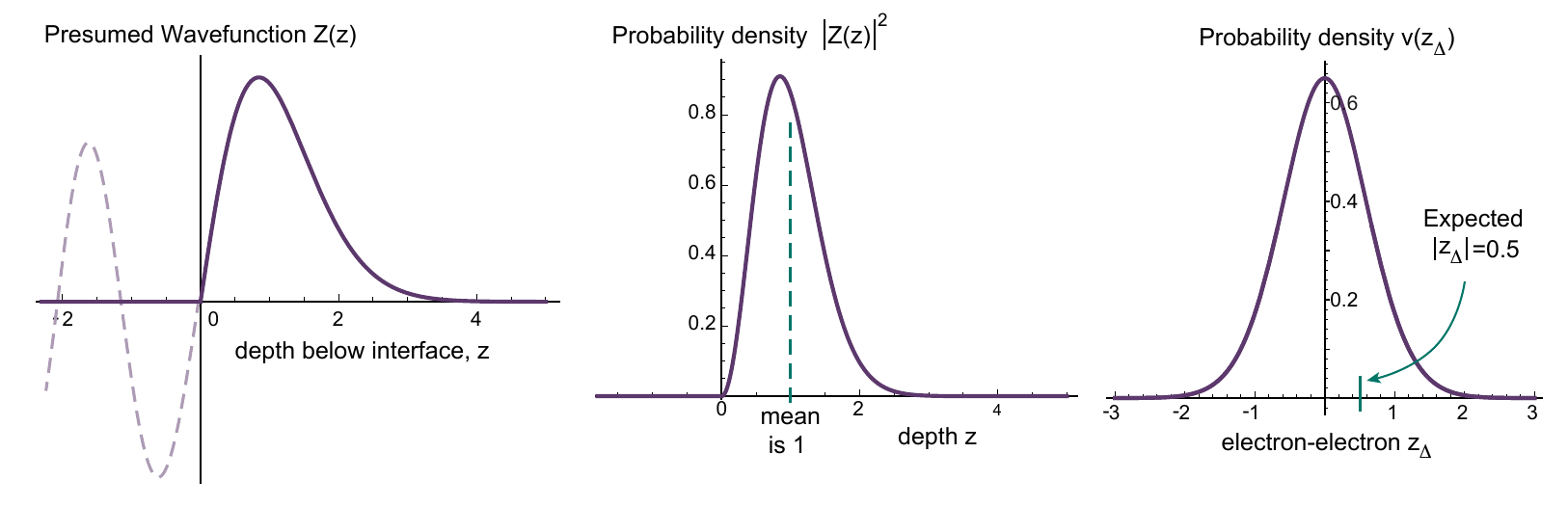}
    \caption{Illustration of the use Airy function in modelling a 3D structure. Left: A single electron's wavefunction has the $z$-dependence shown by the purple line; it is zero for $z\leq 0$, and for positive $z$ it is an Airy function of the first kind. The function is shifted so that its first root is at $z=0$ and, as a useful reference, scaled so that the mean value of $\langle z \rangle=1$ (as shown in the centre panel). The right panel shows the resulting probability distribution for the relative $z$ coordinate between two such electrons; it is similar to a Gaussian and the expected value of $|z_\Delta|$ is $0.5$.}
    \label{fig:Airy}
\end{figure*}

\begin{figure}
    \centering
    \includegraphics[width=\linewidth]{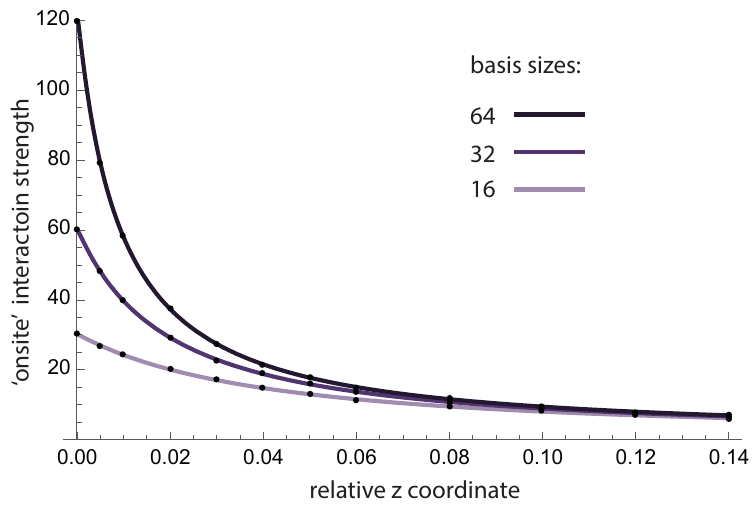}
    \caption{The calculated strengths of $1/\sqrt{\Delta^2+z_\Delta^2}$ where $\Delta$ is the separation of the electrons in the $x$-$y$ plane, in the `shared grid point' case. Key values are also shown in Table\,\ref{tab:coulombTerms}. Lines are fitted power series.}
    \label{fig:zStrengths}
\end{figure}

\begin{table}[]
    \centering
    \begin{tabular}{c|c c c c c c c c}
        \hline
        basis & \ & \ & \ & $z_\Delta$ \\
        size & 0 & 0.005 & 0.01 & 0.02 & 0.03 & 0.04& 0.05 & 0.06 \\
        \hline
       16 & 30.2 & 26.9 & 24.2 & 20.0 &  17.0 & 14.7 & 13.0 & 11.5 \\
       32 & 60.0 & 48.1 & 39.8 & 29.2 &  22.8 & 18.7 & 15.8 & 13.7 \\
       64 &  120.0 & 79.3 & 58.1 &  37.3 &  27.3 & 21.5 & 17.7 & 15.1 \\
       \hline
       \hline
    \end{tabular}
    \caption{Calculated strengths of the term $1/\sqrt{\Delta^2+z_\Delta^2}$, where $\Delta$ is the separation of the electrons in the $x$-$y$ plane, in the `shared grid point' case as defined in the main text. For 2D dots the relevant column is $z_\Delta=0$. Other columns correspond to an enforced $z_\Delta$ as discussed in the section on 3D dots. See also Fig.\,\ref{fig:zStrengths}.  }. 
    \label{tab:coulombTerms}
\end{table}

We further assume that the function $Z(z)$ is part of an Airy function of the first kind, usually written $\text{Ai}(z)$ (see Fig.~(\ref{fig:Airy})). Specifically, 
\begin{eqnarray}
Z(z)&=&N\,\text{Ai}(\frac{z-a_0}{L})\ \ \text{for}\ \ z>L\,a_0\nonumber\\
&=&0\ \ \text{otherwise.}
\end{eqnarray}
Here $a_0\approx -2.3381$ is the first zero of the Airy function, $\text{Ai}(a_0)=0$.
The constant $N$ is simply the normalisation, and $L$ is a constant determining how deep the $z$-axis distribution is.
The treatment of this 3D scenario follows the 2D method closely, and \cref{eqn:periodicC} for the periodic Coulomb interaction is still relevant; we do not need to introduce a periodicity in the $z$-direction since the model has no such periodicity. 
It is convenient to still use $\bm r_1$ 
and $\bm r_2$, and their sum and 
difference $\bm S$ and $\bm \Delta$, where these vectors are understood to line in the $x$-$y$ plane. 
We additionally introduce $z_1$ and $z_2$ for the new dimension. 
Making the natural change of variables $
z_S=z_1+z_2$, $z_\Delta=z_1-z_2$ and noting that the Coulomb interaction has no dependence on $z_S$, we find the relevant function is the probability density $v(z_\Delta)$ where
where
\[
v(z_\Delta)=\frac{1}{2} \int\displaylimits_{-\infty}^\infty Z(\frac{z_S+z_\Delta}{2})^2\ Z(\frac{z_S-z_\Delta}{2})^2 \ dz_S. 
\]
noting $
\int_{-\infty}^\infty v(z_\Delta) dz_\Delta = 1
$.
Then in place of the earlier \cref{eqn:DeltaAlone} we now have
\begin{eqnarray}
&&\bra{n^\prime_{1,x} n^\prime_{1,y} n^\prime_{2,x} n^\prime_{2,y}}H_\text{coul}\ket{0\ 0\ n_{2,x} n_{2,y}}= \qquad\qquad\\
&&\qquad\qquad\qquad
\frac{1}{4}
\int\displaylimits_{-\infty}^\infty\!dz_\Delta\int\displaylimits_\textsc{box}\!\!\! d\bm \Delta\ 
C(|z_\Delta\bm k + \bm \Delta|)\,f(\bm \Delta)\,v(z_\Delta)\nonumber
\label{eqn:3DDeltaAlone}
\end{eqnarray}

This leads ultimately to a 3D generalisation of the earlier \cref{eqn:finalMatrixEle} as follows:
\begin{eqnarray}
&&\bra{n^\prime_{1,x} n^\prime_{1,y} n^\prime_{2,x} n^\prime_{2,y}}H_\text{coul}\ket{0\,0\,n_{2,x} n_{2,y}}=\quad \nonumber\\
&&K\frac{e^{i\Theta}}{4M^4}
\int\displaylimits_{-\infty}^\infty\!
dz_\Delta\, v(z_\Delta)
\int\displaylimits_0^{2\pi}\!\! d\theta \int\displaylimits_0^{\Delta_\text{max}(\theta)}
\!\!\!\!\!\! \frac{\Delta\,d\Delta}{\sqrt{z_\Delta^2+\Delta^2}}\\ 
&&\times h_{n^\prime_{1,x}\,n^{\prime\prime}_{2,x}}(\Delta\cos\theta+\frac{n_{2,x}}{M}) h_{n^\prime_{1,y}\,n^{\prime\prime}_{2,y}}(\Delta\sin\theta+\frac{n_{2,y}}{M})\nonumber
\label{eqn:3DfinalMatrixEle}
\end{eqnarray}
Thus we introduce an offset $z_\Delta$ into the Coulomb interaction, and integrate over $z_\Delta$ with the probability density $v(z_\Delta)$. This can be performed numerically; while the need to integrate over $z_\Delta$ introduces a multiplicative time cost as compared to the 2D computation, the function $v(z_\Delta)$ is a simply Gaussian-like form and therefore requires only a relatively modest sampling resolution. 

The result is to `soften' the Coulomb interaction in the sense that there will be appreciable reduction in the on-diagonal matrix elements for small $n{2,x}$, $n_{2,y}$, especially the $n{2,x}=n_{2,y}=0$ case where the two electrons lie at the same grid point. This is intuitive since the additional dimension affords the electrons `more freedom to avoid one another', so to speak. 

As shown in the main paper, the effect on the qubit-qubit interaction of adopting the 3D potential is to appreciably increase this coupling.

\subsection{Off-diagonal terms}
The expressions derived above allow us to evaluate both the on-diagonal and off-diagonal matrix elements of $H_\text{coul}$ in the real-space Dirichlet basis. This matrix is not strictly diagonal except as the number of basis states $M\rightarrow\infty$ \footnote{See however the approach in Ref.~\cite{RyanLDQSM},  where it is explained that indeed the Coulomb matrix is diagonal if one performs integrals only at the grid points, and this is a valid approach in that it will converge successfully with increasing $M$.}. 

The computations in the main paper simply neglect the off-diagonal terms. Computing their magnitude using the expressions above, one finds for example that in the 3D version and with $M=64$ basis states, the  off-diagonal elements with the greatest amplitude are weaker by more than an order of magnitude than the corresponding on-diagonal element. Moreover, the matrix elements are vanishing small except near the diagonal (`near' meaning that in \cref{eqn:3DfinalMatrixEle}, $n^\prime_{1,x}$, $n^\prime_{1,y}$, $n^{\prime\prime}_{2,x}$ and $n^{\prime\prime}_{2,y}$ are all small or zero).

In a future work it would be interesting to keep (at least) the most significant of the off-diagonal terms present in order to verify that the impact on the quantities of interest is negligible. One could of course modify the dynamical simulation code in various ways to do this, but it might be interesting to apply the method called `augmented split operator' described in Ref.~\cite{hans_grid_based_2023}. This approach was actually suggested as an efficient method of including off-diagonal elements when the simulation itself is running on some future, fault-tolerant quantum computer; however the method should translate well to the present classical code since it is ultimately just a protocol for efficiently manipulating a subset of the amplitudes.

\end{document}